\documentclass{article}

\usepackage{arxiv}
\usepackage[utf8]{inputenc} 
\usepackage[T1]{fontenc}    
\usepackage{hyperref}       
\usepackage{url}            
\usepackage{booktabs}       
\usepackage{amsfonts}       
\usepackage{nicefrac}       
\usepackage{microtype}      
\usepackage{lipsum}
\usepackage[linesnumbered,boxed,ruled]{algorithm2e}
\usepackage{graphicx, subfigure}
\usepackage{array}
\usepackage[thinlines]{easytable}
\usepackage{amsmath,bm}
\usepackage{stmaryrd}
\usepackage{multirow}
\usepackage{mathrsfs}
\usepackage{dsfont}
\usepackage{color}
\usepackage{amsthm}
\usepackage{enumitem}
\usepackage{tikz}
\usepackage{fancybox}

\newcommand{\nosection}[1]{\vspace{2pt}\noindent\textbf{#1.}}

\theoremstyle{definition}
\newtheorem{definition}{Definition}[section]

\title{Survey and Open Problems in Privacy Preserving Knowledge Graph: Merging, Query, Representation, Completion and Applications}

\author{
  Chaochao Chen$^{1}$, Jamie Cui$^{1}$, Guanfeng Liu$^{2}$, Jia Wu$^{2}$, Li Wang$^{1}$\\
  $^{1}$Ant Group, Hangzhou, China \\
  $^{2}$Department of Computing, Macquarie University, Sydney, Australia\\
  \{chaochao.ccc, shanzhu.cjm, aymond.wangl\}@antgroup.com \\
  \{guanfeng.liu, jia.wu\}@mq.edu.au
}

\begin{document}
\maketitle

\begin{abstract}
Knowledge Graph (KG) has attracted more and more companies' attention for its ability to connect different types of data in meaningful ways and support rich data services. However, the data isolation problem limits the performance of KG and prevents its further development. That is, multiple parties have their own KGs but they cannot share with each other due to regulation or competition reasons. 
Therefore, how to conduct \textit{privacy preserving KG} becomes an important research question to answer. That is, multiple parties conduct KG related tasks collaboratively on the basis of protecting the privacy of multiple KGs. 
To date, there is few work on solving the above \textit{KG isolation problem}. In this paper, to fill this gap, we summarize the open problems for privacy preserving KG in data isolation setting and propose possible solutions for them. Specifically, we summarize the open problems in privacy preserving KG from four aspects, i.e., \textit{merging}, \textit{query}, \textit{representation}, and \textit{completion}. We present these problems in details and propose possible technical solutions for them. Moreover, we present three privacy preserving KG-aware \textit{applications} and simply describe how can our proposed techniques be applied into these applications. 
\end{abstract}


\section{Introduction}\label{sec:intro}

Recently, Knowledge Graph (KG) has been popularly constructed and used by more and more companies due to its ability of connecting different types of data in meaningful ways and supporting rich data services. 
A KG is a heterogeneous graph composed of entities (nodes) and relations (edges), and 
in some KGs \cite{besta2019demystifying} there are also properties (features) and labels\footnote{Labels can be taken as special properties, and we use properties to denote both features and labels in the following of this paper.} for entities. A knowledge edge is represented as a factual triple of the form (\textit{head entity}, \texttt{relation}, \textit{tail entity}), also denoted as ($h$, $r$, $t$). 
For example, (\textit{Andrew C. Yao [Scientist,theorist, and Professor]}; \texttt{WinnerOf}; \textit{Turing Award [Alan Turing and ACM]}) is a fact in KG with entity properties. 
So far, KGs have been applied into various tasks such as question answering \cite{zhang2018variational,huang2019knowledge}, recommender system \cite{cao2019unifying,wang2019kgat}, and information extraction \cite{hoffmann2011knowledge,koncel2019text}. Recent advances in KGs include knowledge representation learning \cite{bordes2011learning,nickel2016holographic,wang2017knowledge}, knowledge acquisition and completion \cite{han2018neural,chen2018variational,omran2019embedding}, and knowledge-aware applications \cite{petroni2019language}. 

\begin{figure}
\centering
\includegraphics[width=10cm]{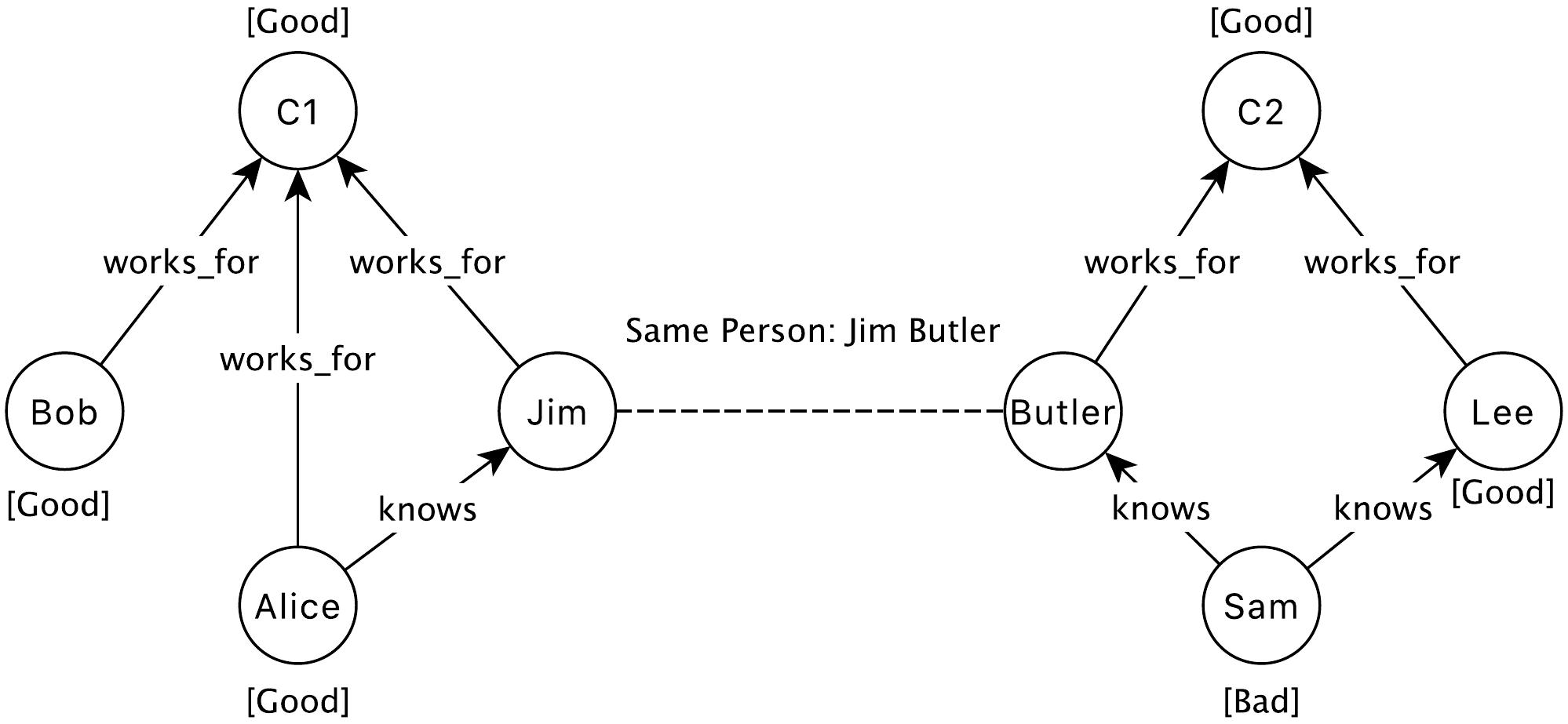}
\caption{A typical case of the KG isolation problem. Party $\mathcal{A}$ (left graph) and party $\mathcal{B}$ (right graph) cannot figure out that $\mathsf{Jim}$ works for both $\mathsf{C1}$ and $\mathsf{C2}$. Besides, party $\mathcal{A}$ is likely to misjudge $\mathsf{Jim}$ as `good' based on its partial KG.}
\label{fig:problem}
\end{figure}

\nosection{KG isolation problem --- an example}
Data isolation has been a long-standing problem ever since, especially with kinds of regulations coming into force all over the world in recent years. 
KG isolation is a typical example of data isolation problem. That is, KGs are isolated by multiple parties, as the example shown in Figure \ref{fig:problem}. 
KG isolation is quite a common problem in practice, since different institutions (e.g., banks, financial companies, and social media platforms) may construct their own KGs with their own data. 
Figure \ref{fig:problem} shows a typical case of the KG isolation problem, where there are two parties and each of them has a KG itself. More specifically, Party $\mathcal{A}$ has four entities $(\mathsf{C1, Bob, Alice, Jim})$ and their relations, and party $\mathcal{B}$ has four entities $(\mathsf{C2, Lee, Butler, Sam})$ and their relations. Besides, each entity has its properties another property, e.g., `good' and `bad'. Due to the data isolation problem, the KGs of party $\mathcal{A}$ and party $\mathcal{B}$ cannot share with each other. Thus, information are also limited to both parties to deploy further artificial intelligence applications. For example, although they have overlap entities ($\mathsf{Jim~ Butler}$), it is difficult for party $\mathcal{A}$ or party $\mathcal{B}$ to figure out the fact that $\mathsf{Jim}$ works for both companies $\mathsf{C1}$ and $\mathsf{C2}$. Nevertheless, when both parties train a model based on their standalone KG, the model of party $\mathcal{A}$ is more likely to misjudge $\mathsf{Jim}$ as $\mathsf{[Good]}$ since a good entity `Alice' knows him, and he works for a good company. However this is not the case in party $\mathcal{B}$, party $\mathcal{B}$ knows that $\mathsf{Jim}$ has relation with a `really bad' people $\mathsf{Sam}$, thus he should be probably labelled as $[\mathsf{Bad}]$.

To solve the data isolation problem, existing works propose different privacy preserving machine learning techniques such as collaborative learning \cite{chase2017private,li2020homopai}, federated learning \cite{yang2019federated,kairouz2019advances}, split learning \cite{vepakomma2018split}, and secure machine learning \cite{mohassel2017secureml,riazi2018chameleon}. 
To date, existing privacy preserving machine learning techniques have covered most traditional data mining and machine learning algorithms, e.g., k-means \cite{mohassel2019practical}, PCA \cite{liu2020privacy}, logistic regression \cite{chen2020homomorphic}, tree based model \cite{fang2020hybrid}, neural network \cite{zheng2020industrial}, and recommender system \cite{chen2018privacy,chen2020practical}. 

So far, there have been several literatures on privacy preserving graph algorithms \cite{brickell2005privacy,he2011privacy,blanton2013data,sharma2016privacy,chang2016privacy}. For example, Brickell and Shmatikov \cite{brickell2005privacy} proposed secure methods for finding shortest distance on multiple graphs. 
Graph anonymization and private link discovery approaches were presented in \cite{he2011privacy}, and data-oblivious graph analysis algorithms were provided in \cite{blanton2013data}. Besides, the authors in \cite{sharma2016privacy,chang2016privacy} proposed secure methods for graph analysis on encrypted graphs in cloud computing setting. 
However, how to perform privacy preserving KG under KG isolation setting remains an open research problem.
This is potentially because two reasons. On the one hand, data in KG not only contains entities (samples) and properties (features), but also involves different kinds of relations between entities, which is more complicated than data in traditional machine learning. On the other hand, techniques in KG usually involve machine learning approaches such as deep learning and thus are more complex than naive graph analysis algorithms. 

In this paper, to fill this gap, we aim to summarise the open problems for privacy preserving KG in data isolation setting. That is, there are multiple parties and each of them has a KG constructed by its own private data The purpose of privacy preserving KG is to perform KG related tasks using the KGs from multiple parties, on the basis of protecting the data privacy, meanwhile achieving comparable performance as the plaintext KG by merging the raw KGs of multiple parties. 

\begin{figure*}
\centering
\includegraphics[width=15.5cm]{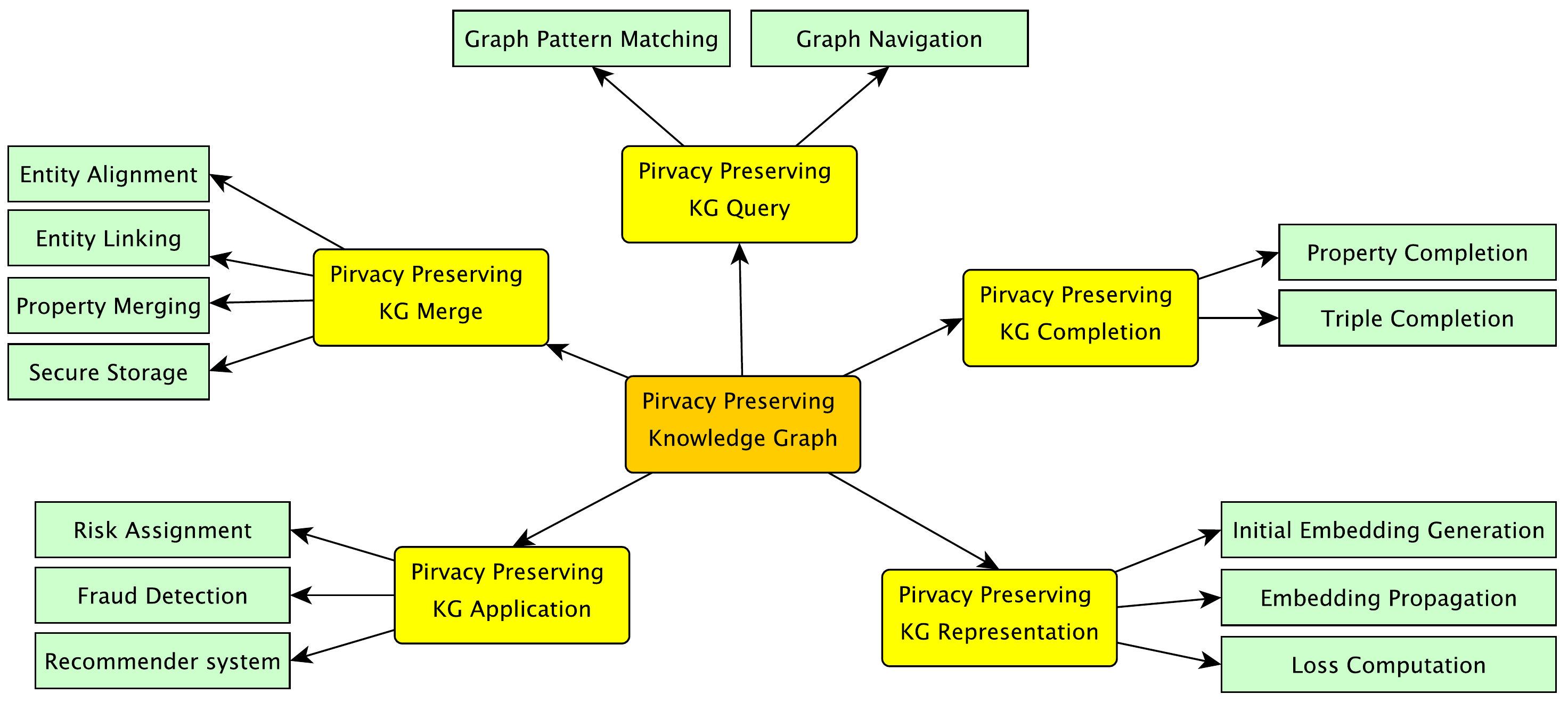}
\caption{Open problems in privacy preserving knowledge graph.}
\label{fig:overview}
\end{figure*}

\nosection{Open problems in privacy preserving KG}
Motivated by the existing advanced techniques in graph learning and KG, in this paper, we summarize the open problems in privacy preserving KG from five aspects, i.e., merging, query, representation, completion, and applications, as is shown in Figure \ref{fig:overview}.

\textit{Privacy preserving KG merging.}
KG isolation problem describes the facts that parties independently holds their only private KGs, therefore the first open problem for privacy preserving KG is to merge these KGs and store the merged result securely. 
For merging, different parties are likely to have different entity sets, and therefore, the most important thing is to identify their common entities and merge the corresponding properties privately. A possible solution is private set intersection \cite{pinkas2018scalable} and secure Multi-Party Computation (MPC) \cite{yao1986generate,lindell2020secure}. 
For storage, the troublesome problem is how to keep the merged result secure and maintain doable for the following operations and application on KG. Homomorphic encryption \cite{gentry2009fully} and secret sharing \cite{shamir1979share} are possible ways for this.

\textit{Privacy preserving KG query.}
For traditional KG, query is quite straightforward for a single party by using mature graph traversal languages such as Gremlin \cite{rodriguez2015gremlin}.
However, under privacy preserving KG setting, query is challenging considering the privacy constrain. That is, the one who initiates the query cannot obtain any other information except the query result, and the KG parties do not know what the query is. Intuitively, this can be done directly by using Oblivious Transfer (OT) \cite{rabin2005exchange} or private information retrieval \cite{chor1995private}. 
Unfortunately, privacy preserving KG query is more challenging since these KGs are isolated by multi-parties, and therefore, it is the second open problems in privacy preserving KG. 
A possible solution for this is combining OT with other cryptographic techniques such as secret sharing \cite{shamir1979share},  garbled circuit \cite{yao1986generate}, and Pseudo-Random Generator (PRG) \cite{haastad1999pseudorandom}. 

\textit{Privacy preserving KG representation.}
KG representation learning is a critical research direction of KG, which aims to learn low-dimensional embeddings of entities and relations, paving the way for many knowledge completion tasks and downstream applications \cite{ji2020survey}. 
In KG isolation setting, after KG merging, data (including entities, properties, and relations) are securely stored by multi-parties. Although existing secure machine learning as a service \cite{mohassel2017secureml} provides methods for representing entities (samples) and properties (features), it cannot capture the relations between entities, and thus cannot achieve comparable performance as the traditional plaintext KG representation learning approaches. 
Therefore, privacy preserving KG representation is the third open problem in privacy preserving KG. A possible solution could be combining graph neural (convolution) network \cite{cai2018comprehensive,liu2019geniepath} with the above mentioned cryptographic techniques to build privacy preserving graph neural (convolution) network. The learned representations can be stored in plaintext or encrypted format based on privacy requirement. 

\textit{Privacy preserving KG completion.}
KG completion (aka reasoning) is an active field of research since KGs are known for their incompleteness and noise. Existing researches on KG completion include entity property prediction \cite{lin2015learning}, and triple classification \cite{dong2019triple}. With the learned KG representations for entities and relations, in either plaintext or encrypted format, one can build secure machine learning algorithms for further KG completion tasks. The challenge here is how to provide scalable and flexible secure machine learning frameworks so that one can easily build secure machine learning algorithms to meet the various needs in KG completion.
One possible solution is a system with hybrid secure computation protocols, rich computation operations, and powerful domain specific language.  

\textit{Privacy preserving KG-aware applications.} KG and its related techniques have boosted the performance of numerous applications such as natural language understanding \cite{logan2019barack}, question answering \cite{chen2019bidirectional}, fraud detection \cite{liu2018heterogeneous}, risk assessment \cite{cheng2019risk}, and recommender system \cite{wang2019kgat}. Although one cannot enumerate all the possible KG-aware applications in KG isolation setting, we showcase three real-world privacy preserving KG-aware applications and present how to solve them using the four privacy preserving KG techniques above. 

The following paper is organized as follows. 
Section \ref{sec:pre} describes definition, related secure computation techniques, and threat model. Section \ref{sec:merging}-Section \ref{sec:app} present the detailed open problems and possible solutions in privacy preserving KG, i.e., merging, query, representation, completion, and applications. Section \ref{sec:conclusion} concludes this paper. 
\section{Definitions and Backgrounds}\label{sec:pre}

We first give the definitions for traditional KG and privacy preserving KG. We then describe background knowledge on secure computation techniques. 
We finally describe the threat models in privacy-preserving KG, which models the adversaries' behaviour.
Table 1 lists the main notations used throughout the paper.

\begin{table}
\centering
\caption{Summary of Notations.}
\begin{tabular}{cc}
  \toprule
  Notation & Explanation \\
  \midrule
  $\mathcal{G}=\{\mathcal{E}, \mathcal{R}, \mathcal{F}\}$ & Knowledge graph (KG) \\
  $\mathcal{E}$ & Entity set \\
  $\mathcal{R}$ & Relation set \\
  $\mathcal{F}$ & Fact set \\
  $h$ / $t$ & Head/tail entity \\
  $\mathcal{X}$ & Property set \\
  $x \in \mathcal{X}$ & A property (feature) \\
  $U, U^\prime$ & Location set in KG \\
  $\Psi$ & Query instruction set \\
  $r \in \mathcal{R}$ & A relation between entities \\
  $\langle x \rangle$ & Secret sharing of $x$ \\
  $\textbf{h}$ & Entity embedding \\
  $\textbf{W}$ & Weight matrix \\
  $\sigma$ & Non-linear active function \\
  $\gamma$ & Traversal condition \\
  $q$ & Polynomial coefficients \\
  $\mathcal{N}$ & Neighborhood function \\
  $\text{AGG}_k$ & Aggregator of $k$-th depth propagation \\
  $K$ & Propagation depth \\
  $\mathcal{A}$, $\mathcal{B}$ & Parties who own KG \\
  \bottomrule 
\end{tabular}
\label{tab:natations}
\end{table}

\subsection{Definition} 

\textit{Definition of traditional KG.}
Following previous literature \cite{wang2017knowledge,ji2020survey}, we define a knowledge graph as $\mathcal{G}=\{\mathcal{E}, \mathcal{R}, \mathcal{F}\}$, where $\mathcal{E}$, $\mathcal{R}$ and $\mathcal{F}$ are sets of entities, relations, and facts, respectively. A fact is denoted as a triple $(h, r, t) \in \mathcal{F}$ which is composed of a head entity $h \in \mathcal{E}$, a tail entity $t \in \mathcal{E}$, and a relation $r \in \mathcal{R}$ between them, e.g., (\textit{Andrew C. Yao}; \texttt{WinnerOf}; \textit{Turing Award}). Besides, entities may have a property set $\mathcal{X}$ that describe them. For each entity, there is a property sub-set $x \in \mathcal{X}$ that describes it, e.g., \textit{Andrew C. Yao [Scientist, theorist, and Professor]} and \textit{Turing Award [Alan Turing and ACM]}.

\textit{Definition of privacy preserving KG. }
Assume there are $n$ parties and each of them has an individual KG $\mathcal{G}_i=\{\mathcal{E}_i, \mathcal{R}_i, \mathcal{F}_i\}$, where $i\in \{1, 2, ..., n\}$. 
Same as the traditional KG, the fact set of party $i$ contains the triples $(h_i, r_i, t_i) \in \mathcal{F}_i$ which contains a head entity $h_i \in \mathcal{E}_i$, a tail entity $t_i \in \mathcal{E}_i$, and a relation $r_i \in \mathcal{R}_i$ between them. 
Besides, for each party $i$, there is a property set $\mathcal{X}_i$ describing its entities. 
The purpose of privacy-preserving KG is to conduct KG-related tasks (including query, representation, completion, and application) on the basis of (1) preserving the data confidentiality of the KGs held by $n$ parties, and (2) achieving comparable performance as the traditional KG on mixed plaintext data. 

\subsection{Secure Computation}\label{sec:pre-sc} 

Secure computation is a general cryptography term, encompassing all methods that allows computation on data while still keeping data private. 

It is also considered as the core technique for implementing a privacy-preserving application. In the literature, secure computation has directed research into generic solutions such as \emph{Homomorphic Encryption} (HE) \cite{Gentry2010ComputingAF,Fan2012SomewhatPF}, \emph{Oblivious Random Access Memory} (ORAM) \cite{Goldreich1987TowardsAT,Goodrich2011ObliviousRS} and \emph{Universal Circuit} (UC) \cite{Kolesnikov2008APU, Lipmaa2016ValiantsUC,Gnther2017MoreEU}, and also inspired works on primitives with a specific question, such as \emph{Oblivious Transfer} (OT) \cite{Ishai2003ExtendingOT,Asharov2013MoreEO, Chou2015TheSP}, \emph{Private Information Retrieval} (PIR) \cite{Kiayias2015OptimalRP, Canetti2017TowardsDE, Boyle2017CanWA, Patel2018PrivateSI, Angel2018PIRWC, Ali2019CommunicationComputationTI} and \emph{Private Set Intersection} (PSI) \cite{Chase2020PrivateSI}.

Though the problem of secure computation has been studied for almost 30 years, most of the works are theoretical. Recently, the rapid development of computer networking has pushed one solution into practical --- secure multi-party computation (MPC) \cite{yao1986generate,lindell2020secure}. In this work, we will take MPC as a typical technique to solve privacy preserving KG problems. More specifically, we will focus on secret sharing technique. 

\nosection{Secret sharing (SS) \cite{shamir1979share}} Secret Sharing is an essential cryptographic primitive for many MPC protocols. Roughly, a secret sharing scheme splits a secret value into multiple pieces, such that the secret is only revealed with sufficient number of pieces. Formally, a secret sharing scheme comprises two algorithms $(\mathsf{Shr}, \mathsf{Rec})$, and we use a pair with angle brackets, i.e., $\langle x \rangle$, to denote $x$ is secret shared. 
Take a two party ($\mathcal{A}$ and $\mathcal{B}$) secret sharing for example,
considering that party $\mathcal{A}$ wants to share ($\mathsf{Shr}$) its private data $x$ with party $\mathcal{B}$, $\mathcal{A}$ first randomly generates a share $\langle x \rangle _0 \in \mathds{Z}_{P}$ with $P$ denoting a large prime, and keeps $\langle x \rangle _0$ itself. Then $\mathcal{A}$ calculates $\left\langle x \right\rangle_1 = x- \langle x \rangle _0$ mod $P$, and sends $\left\langle x \right\rangle_1$ to $\mathcal{B}$. 
To reconstruct ($\mathsf{Rec}$) data $x$, which is shared between both parties, one party obtains the share from the other party, and then calculates $x=\langle x \rangle _0 + \langle x \rangle _1$ mod $P$. 




We now list secret sharing based secure computation primitives used in this paper as follows. 

\begin{itemize} [leftmargin=*] \setlength{\itemsep}{-\itemsep}
\item \textbf{LINEAR}
$\langle c \rangle \leftarrow \alpha \cdot \langle a \rangle + \langle b \rangle + \beta$ for secretly shared values $\langle a \rangle$, $\langle b \rangle$, and plaintext values $\alpha$ and $\beta$. Linear operations can be done by each party locally without interacting with other parties. And the result $\langle c \rangle$ is also shared. 

\item \textbf{MUL} 
$\langle c \rangle \leftarrow \langle a \rangle \cdot \langle b \rangle$ for secretly shared values $\langle a \rangle$ and $\langle b \rangle$, such that $c = a \cdot b$ and returns a shared value $\langle c \rangle$. Secretly shared multiplication replies on Beaver's Triple technique \cite{beaver1991efficient} which needs interactions between participants. 

\item \textbf{DIV}
$\langle c \rangle \leftarrow \langle a \rangle / \langle b \rangle$ for secretly shared values $\langle a \rangle$ and $\langle b \rangle$, such that $c = a / b$ and returns a shared value $\langle c \rangle$. Secretly shared multiplication can be implemented using numerical optimization algorithms such as Goldschmidt’s series expansion algorithm \cite{goldschmidt1964applications}. After it, \textbf{DIV} can be approximated and computed by \textbf{LINEAR} and \textbf{MUL}.

\item \textbf{ARGMAX}
$\arg\max\limits_{i}(\langle a_i \rangle) \leftarrow (\langle a_1 \rangle, \langle a_2 \rangle, ..., \langle a_I \rangle)$ for a list of shared values and returns the one with the maximum value. This can be done by conducting $I-1$ secure comparison using Boolean secret sharing \cite{demmler2015aby}. It can also speedup by using tree structure based parallel comparison \cite{mohassel2019practical}. 

\end{itemize}

Note that secret sharing only works in finite field to guarantee security, and fixed-point representation is popularly used to make it suitable for float numbers \cite{mohassel2017secureml}. 

\subsection{Threat Model}
The threat model of PPKG follows the standard MPC security under the real world vs. ideal world paradigm \cite{lindell2020secure}. That is to say, we categorize the adversary's behaviours into one of the following:

\begin{itemize}
    \item \emph{Semi-honest adversary} who corrupts parties but follows the protocol as it specified. 
    \item \emph{Malicious adversary} who causes corrupted parties to deviate arbitrarily from the prescribed protocol in an attempt to violate security.
\end{itemize}

\section{Privacy Preserving KG Merging}\label{sec:merging}
In this section, we first describe the problems in privacy preserving KG merging and our proposed secure merging solutions. We then present how to store the merged KG securely based on secret sharing.

\begin{figure}
\centering
\includegraphics[width=10cm]{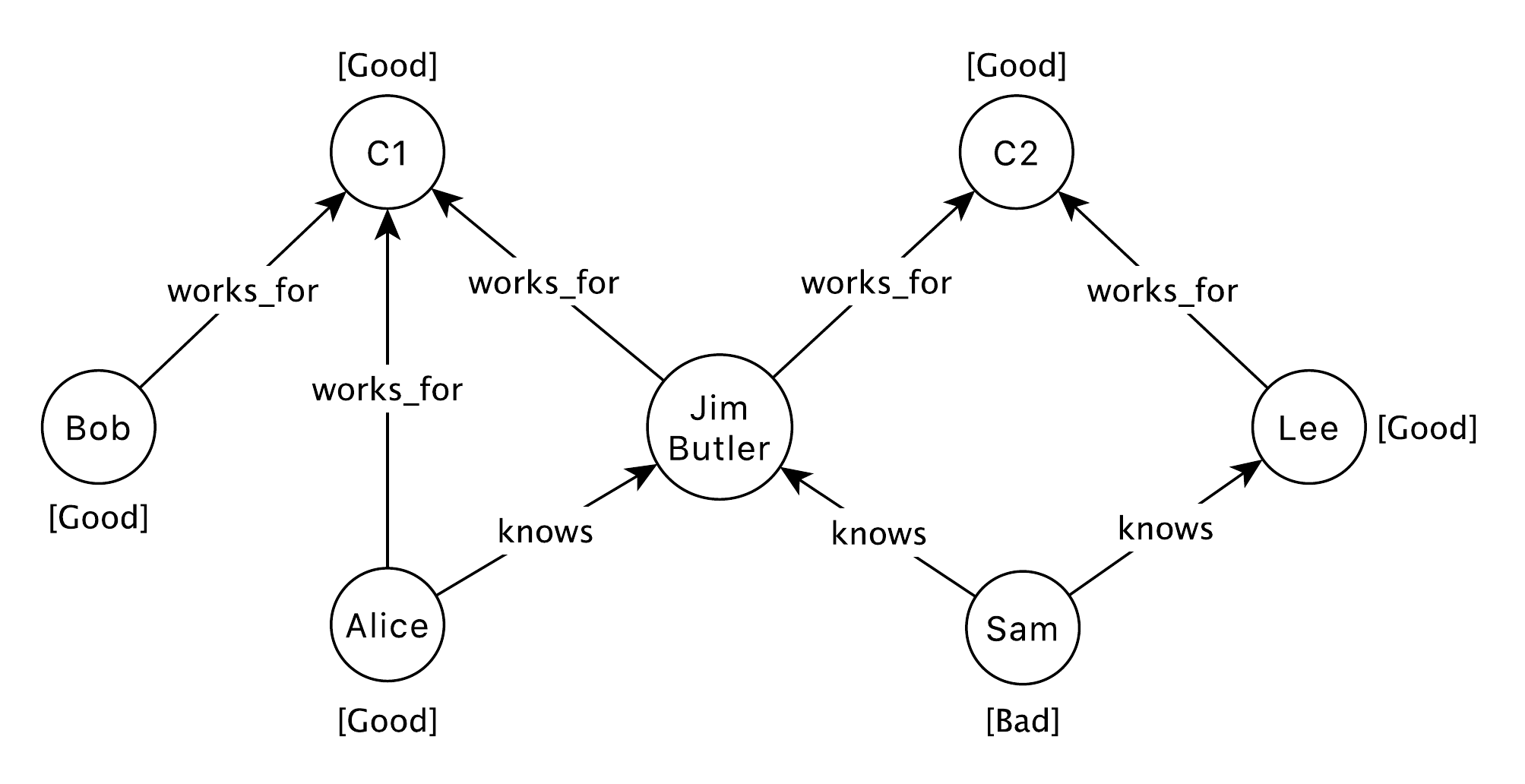}
\caption{Merged KG from Figure \ref{fig:problem}. }
\label{fig:merging}
\end{figure}

\subsection{Merging}
Under KG isolation setting, KGs are separated by multiple parties, and naturally, the first step for privacy preserving KG is to merge these KGs and store the merged result securely. 
Privacy preserving KG merging has several challenging problems. 
First, multi-parties have their own entity sets, and how to align their entities privately is the first challenging problem. Second, the KGs of different parties are built from different data sources, hence, their entity names may be different from each other although those names have the same meaning. For instance, Party $\mathcal{A}$ may mark the node as ``$\mathsf{Jim}$'', while Party $\mathcal{B}$ mark the node as ``$\mathsf{Jimmy}$'', however both of which refer to the same entity. Therefore, how to conduct entity linking across different parties privately is the second problem. 
Third, the same entity across multi-KGs may also have different property values. For example, the KG of party $\mathcal{A}$ shows $\mathsf{Jim}$ has a property of 25 years old, while the KG of party $\mathcal{A}$ indicates $\mathsf{Jim}$ is 35 years old. 
Therefore, how to merge the property values for the same entity becomes another problem. We now describe these challenging problems in details and present possible solutions. 

\subsubsection{Private Entity Alignment}
Under KG isolation setting, private entity alignment aims to align the same entities among different KGs. 
Formally, it is defined as follows. 

\begin{definition}[Private entity alignment]
Given any two KGs, namely $\mathcal{G}_i=\{\mathcal{E}_i, \mathcal{R}_i, \mathcal{F}_i\}$ and
$\mathcal{G}_j=\{\mathcal{E}_j, \mathcal{R}_j, \mathcal{F}_j\}$, where $i, j \in \{1, 2, ..., n\}$ and $i \ne j$, private entity alignment aims to find the common entity set of $\mathcal{E}_i$ and $\mathcal{E}_j$, and meanwhile protect other private information of both KGs. 
\end{definition}

Take the example in Figure \ref{fig:problem}, private entity alignment will find the common entity set (entity 2 and entity 3) for party $\mathcal{A}$ and party $\mathcal{B}$ and keep other information private. 

\nosection{Possible solutions for private entity alignment}
We find that private entity alignment has exactly the same purpose as one of the existing primitives called PSI \cite{pinkas2018scalable}. 
PSI is a cryptographic protocol that aims to compute the intersection of two sets held by two parties, during which both parties should learn the elements (if any) common to both sets and nothing (or as little as possible) else \cite{de2010practical}. PSI has been extensively researched recently. So far, there are several types of PSI, including public-key based PSI \cite{chen2017fast}, circuit based PSI \cite{huang2012private}, and OT (extension) based PSI \cite{pinkas2018scalable}. One can directly apply these PSI protocols for private entity alignment task. 

\subsubsection{Private Entity Linking} 
Entity linking (disambiguation) is a long-standing problem in KG \cite{moro2014entity}. In KG isolation setting, private entity linking is defined as follows.

\begin{definition}[Private entity linking]
Given any two KGs, $\mathcal{G}_i=\{\mathcal{E}_i, \mathcal{R}_i, \mathcal{F}_i\}$ and
$\mathcal{G}_j=\{\mathcal{E}_j, \mathcal{R}_j, \mathcal{F}_j\}$, where $i, j \in \{1, 2, ..., n\}$ and $i \ne j$, for a target entity $e_i \in \mathcal{E}_i$, private entity linking aims to find the entity $e_j \in \mathcal{E}_j$ that is the matched with $e_i$ (if have), and meanwhile protect the data privacy of both KGs. 
Note that, here $e_i \ne e_j$, since private entity alignment has solved this case. 
\end{definition}

\begin{figure*}
\centering
\includegraphics[width=\columnwidth]{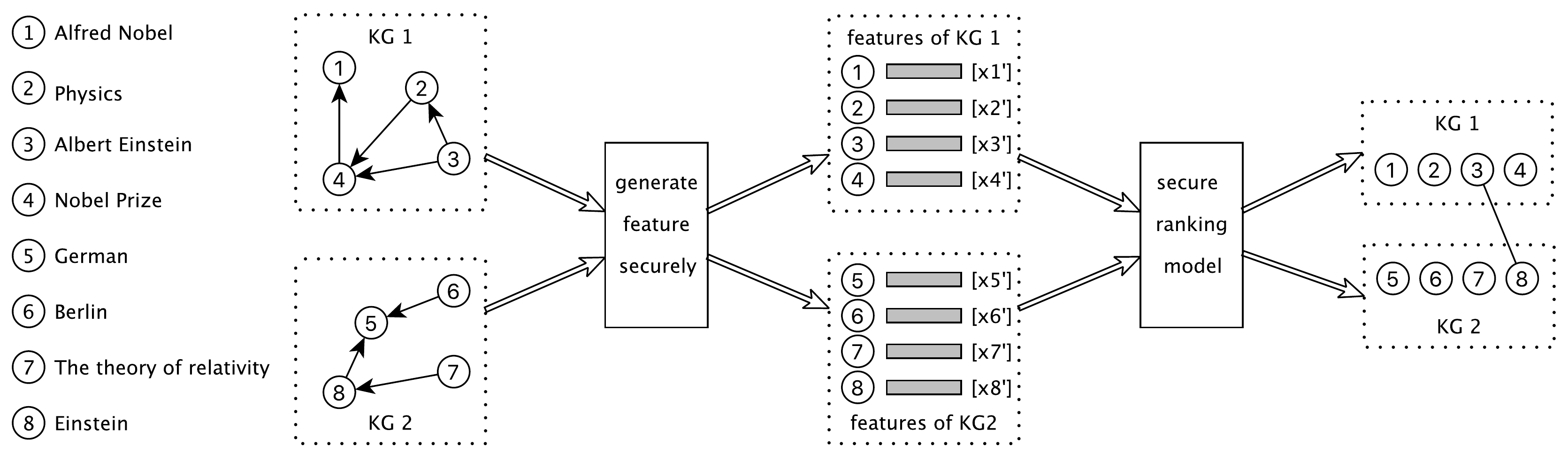}
\caption{Solution of private entity linking. It mainly has two steps, i.e., generate entity features securely based on KGs and ranking entities securely based on features. }
\label{fig:entity-linking}
\end{figure*}

Take Figure \ref{fig:entity-linking} as an example, for the target entity `Albert Einstein' of one party, private entity linking aims to find its matched entity `Einstein' from the other party and meanwhile keep the entities, relations, and descriptions private. 

Existing work on plaintext entity linking mainly have two steps, i.e., candidate entity generation and candidate entity ranking \cite{shen2014entity}. Candidate entity generation can be done by either using \textit{lexical based} method or \textit{semantic based} method. Lexical based methods directly calculate the text similarity of entities and their properties \cite{zhang2010entity}. Later on, semantic based methods, mostly neural models, are popularly used to learn entity embeddings \cite{francis2016capturing,chen2018bilinear,sun2017cross}. 
Candidate entity ranking aims to select the most relevant candidate entity to link for the target entity. The traditional ranking models, e.g., logistic regression, tree based model, and neural networks can be directly used for ranking. 
However, none of them is designed to protect data privacy in KG isolated setting. 

Recently, there are several work on securely calculating the similarity of texts \cite{gondree2009longest,pang2010privacy,reich2019privacy}. For example, Gondree and Mohassel proposed a secure method for calculating the longest common subsequence between two parties \cite{gondree2009longest}. Pang et al. proposed a private text retrieval method using encryption technique \cite{pang2010privacy}. 
Reich et al. proposed secure text classification protocol using MPC \cite{reich2019privacy}. Although their protocol can be applied for private entity linking, but it fails to capture the semantic information between entities. 

\nosection{Possible solutions for private entity linking}
Private entity linking can be taken as a ranking problem, as described above. Therefore, we design a two-step solution for it, as is shown in Figure \ref{fig:entity-linking}, i.e., generating entity features securely based on KGs and ranking entities securely based on features. Note that we do not list candidate entity generation as the first step, but one can do it beforehand to decrease the size of entity feature generation and entity ranking. 

\textit{Generating entity features securely based on KGs.}
Entity features should represent the characteristics (e.g., lexical aspect or semantic aspect) of entities. The purpose of this step is to generate entity features for all the parties with all the KG data, and keep these data privately at the same time. 
We propose to generate entity features for multi-parties securely using whatever information available, e.g., entity name and property. However, there is few existing solutions on generating entity features under privacy-preserving setting, especially semantic based methods. Therefore, there is an urgent need of privacy-preserving semantic models such as word2vec \cite{mikolov2013distributed} and doc2vec \cite{le2014distributed}. 
After generating entity features securely, each party $i$ holds the encrypted or plaintext entity features $[\mathcal{X}'_i]$ or $\mathcal{X}'_i$, based on specific security requirement.

\textit{Ranking entities securely based on features.}
Entity ranking aims to find the best matching entity from the other KG (if have), for a target entity in a given KG, based on the generated entity features. 
We propose to rank entities securely using either unsupervised learning models, e.g., privacy preserving nearest neighbor \cite{shaneck2009privacy}, or supervised learning models, e.g., \cite{fang2020hybrid,chen2020homomorphic}. After this, a party will link its entities to the entities of other parties. As the example shown in Figure \ref{fig:entity-linking}, entity 3 (`Albert Einstein') in KG 1 is linked to entity 8 (`Einstein') in KG 2. 

\subsubsection{Private Property Merging}
After private entity alignment and private entity linking, it will find the \textit{common entity set} between different parties, which includes the same entities (e.g., entity 2 and entity 3 in Figure \ref{fig:problem}) and the linked entities (e.g., entity 3 and entity 8 in Figure \ref{fig:entity-linking}). However, these entities may have different properties, e.g., `physicist' for entity 3 and `actor' for entity 8 in Figure \ref{fig:entity-linking}. The simplest solution for this is taking `physicist' and `actor' as two different properties, but doing this will introduce more noise. Formally, we define private property merging as follows. 

\begin{definition}[Private property merging]
Given the aligned or linked common entity set ($\tilde{\mathcal{E}}$) of $m$ KGs, where $m \le n$, private property merging aims to merge the properties ($\tilde{\mathcal{X}}$) of $\tilde{\mathcal{E}}$ while keep each party's property private. 
\end{definition}

\nosection{Possible solutions for private property merging}
The best solution for private property merging is making secure computations based on different property types, as is shown in Table \ref{tab:property-merging}. 
We divide entity properties into two type, i.e., continuous or discrete properties and categorical properties. The former are real-valued variables (e.g., height) or numeric variables (e.g., age) that have comparison relations, while the later are a limited number of categories or distinct groups (e.g., gender and country) that usually do not have a logical order. We propose different merging operations for these two kinds of properties. First, for continuous and discrete properties, we propose three secure merging operations, i.e., max, min, and (weighted) average. This can be done by using secure MPC techniques, e.g., garbled circuit and secret sharing. An example of this is that, party $\mathcal{A}$ has a property (age, 23) and party $\mathcal{B}$ has a property (age, 15), and one possible merged property is their average (age, 19). Second, for categorical properties, we propose binary or multiple selection, i.e., select the most likely property value from two or multiple values, based on certain rules (e.g., voting). This could be done by using garbled circuits. A typical example is that, for the same entity property (gender), party $\mathcal{A}$ has value `M', party $\mathcal{B}$ has value `F', party $\mathcal{C}$ has value `F', and the merged gender is their majority (`F'). However, these merged result should be stored securely to prevent information leakage. 

\begin{table*}
\centering
\caption{Solution of private property merging.}
\begin{tabular}{cp{4.5cm}<{\centering}p{5.5cm}<{\centering}}
  \toprule
  property type & secure merging operation & example \\
  \midrule
  continuous or discrete & max, min, (weighted) average & $x_1$: 23, $x_2$: 15, merged: 19 \\
  categorical & binary or multiple selection & $x_1$: `M', $x_2$: `F', $x_3$: `F', merged: `F' \\
  \bottomrule 
\end{tabular}
\label{tab:property-merging}
\end{table*}

\subsection{Storage}\label{sec:merge:storage}
The purpose of KG storage is to securely store the KGs after merging for each party. After merging, the relations and properties of the \textit{common entity set} (including aligned entities and linked entities) will naturally change. Therefore, the main challenge here is how to securely store the relations and properties of the common entity set to facilitate subsequent KG tasks, such as query, representation, and completion. 

To date, there are several ways to store KGs, e.g., triple table \cite{harris20033store}, property table \cite{wilkinson2006jena}, and DB2RDF \cite{bornea2013building}. For example, we can represent the property graph model in Figure \ref{fig:problem} in the form of triple table and property table. Given the assumption that each entity have two properties, Table \ref{tab:representation} describes the storage structure of KGs. After merging, each party still has its own triples, however, its properties have been merged with other parties and thus have to be saved in a secure manner to prevent potential information leakage.



\begin{table}
\caption{Representing KGs in Figure \ref{fig:problem} by triple tables and property tables. Party $\mathcal{A}$'s graph is described by the first two tables on the left, party $\mathcal{B}$'s graph is described by the last two tables on the right.}
\centering
\begin{minipage}{0.23\textwidth}
\centering
        \begin{tabular}{@{}l ccc@{}}
            \toprule
            Entity & Relation & Entity \\
            \midrule
            Alice & 1 & C1			\\
            Alice & 2 & Jim 			\\
            Bob & 1 & C1 			\\
            Jim & 1 & C1 			\\
            \bottomrule 				\\
        \end{tabular}
\end{minipage}\hfill
\begin{minipage}{0.235\textwidth}
\centering
        \begin{tabular}{@{}l ccc@{}}
            \toprule
            Entity & Property 	\\
            \midrule
            Alice & (0.8, 1) 	\\
            Bob & (0.5, 1) 		\\
            Jim & (0.8, 1) 		\\
            C1 & (0.7, 1) 		\\
            \bottomrule			\\
        \end{tabular}
\end{minipage}
\begin{minipage}{0.245\textwidth}
\centering
        \begin{tabular}{@{}l ccc@{}}
            \toprule
            Entity & Relation & Entity \\
            \midrule
            Sam & 2 & Lee \\
            Sam & 2 & Butler \\
            Butler & 1 & C2 \\
            Lee & 1 & C2 \\ 			
            \bottomrule 				\\
        \end{tabular}
\end{minipage}
\begin{minipage}{0.245\textwidth}
\centering
        \begin{tabular}{@{}l ccc@{}}
            \toprule
            Entity & Property 	\\
            \midrule
            Sam & (0.1, 1) \\
            Butler & (0.8, 1) \\
            Lee & (0.3, 0) \\
            C2 & (0.4, 1) \\
            \bottomrule			\\
        \end{tabular}
\end{minipage}
\label{tab:representation}
\end{table}

\begin{table}
\begin{minipage}{0.48\textwidth}
\centering
        \caption{$\mathcal{A}$'s property table after merging.}
        \begin{tabular}{@{}l ccc@{}}
            \toprule
            Entity & Property 	\\
            \midrule
            Alice & (0.8, 1) 	\\
            Bob & (0.5, 1) 		\\
            Jim Butler (JB) & ($\langle x^{\mathsf{JB}, 1}\rangle_\mathcal{A}, \langle x^{\mathsf{JB}, 2}\rangle_\mathcal{A}$) \\
            C1 & (0.7, 1) 		\\
            \bottomrule			\\
        \end{tabular}
        \label{tab:APT}
\end{minipage}\hfill
\begin{minipage}{0.48\textwidth}
\centering
        \caption{$\mathcal{B}$'s property table after merging.}
        \begin{tabular}{@{}l ccc@{}}
            \toprule
            Entity & Property 	\\
            \midrule
            Sam & (0.1, 1) \\
            Jim Butler (JB) & ($\langle x^{\mathsf{JB}, 1}\rangle_\mathcal{B}, \langle x^{\mathsf{JB}, 2}\rangle_\mathcal{B}$)\\
            Lee & (0.3, 0) \\
            C2 & (0.4, 1) \\
            \bottomrule			\\
        \end{tabular}
        \label{tab:BPT}
\end{minipage}
\end{table}

\nosection{Possible solutions for secure KG storage}
Two kinds of methods are popularly used to store data securely, i.e., \textit{secret-shared} method and \textit{encrypted} method. 
The former stores data in secret sharing format \cite{lai2019graphse2}. That is, each party holds a share of the raw data, as described in Section \ref{sec:pre-sc}. 
The later stores data in the form of ciphertext \cite{cash2013highly}. That is, each party holds the encrypted data whose secret key is kept by the other party. Following these existing research, we propose to store entity property table securely using either secret-shared method or encrypted method. Table \ref{tab:APT} and Table \ref{tab:BPT} show how to store KGs using secret-shared method, where we use $x^{i,j}$ to denote $j$-th property of entity $i$, for $i \in \{A, B, ..., F\}$ and $j \in \{1, 2\}$.

\section{Privacy Preserving KG Query}
In this section, we first describe the taxonomy of KG queries. Then we describe the problem of privacy-preserving KG query. We finally present the possible solutions for the above problem.

\subsection{Taxonomy of KG Queries}
Essentially, a KG query language is in the form of a regular graph query language but targets at knowledge graphs. Industrial graph databases, such as SPARQL \cite{prudhommeaux2008sparql}, Cypher \cite{francis2018cypher}, and Gremlin \cite{rodriguez2015gremlin}, adopt the property graph model (which is more elaborate) for knowledge graphs. In summary, KG query is a graph query which supports search functionality over property graphs. Angles et al. \cite{Angles2017FoundationsOM} inspected the theory of graph queries in details, and summarized that graph query languages share two most fundamental querying functionalities: \emph{Graph Pattern Matching} (GPM) and \emph{Graph Navigation} (GN).

\nosection{Graph Pattern Matching (GPM)}
The core functionality of answering graph queries is graph pattern matching. Graph pattern is a graph-structured query with variables and constants. For the case of property graph model, the set of $\{\mathcal{E}, \mathcal{R}, \mathcal{F}\}$ is defined as constant, denoted as $\mathsf{Const}$, while the query is defined as the variable set, denoted as $\mathsf{Var}$. For instance, a query ``Search for the people that \texttt{Marko} \texttt{knows}'' towards graph $\mathcal{G} = \{\mathcal{E}, \mathcal{R}, \mathcal{F}\}$ could be turned into a graph pattern $\tilde{\mathcal{G}} = (\tilde{\mathcal{E}}, \tilde{\mathcal{R}}, \tilde{\mathcal{F}})$, where  $\tilde{\mathcal{E}}=\{\texttt{``Marko''}, \delta\}$, and $\tilde{\mathcal{R}}=\{(\texttt{``Marko''}, \texttt{``knows''}, \delta)\}$. To successfully find a match, we first match the graph pattern $\tilde{\mathcal{G}}$ to the graph $\mathcal{G}$, and then search for the occurrences of this pattern. Technically, the match for a KG graph pattern is defined as follows:

\begin{definition}[Match]
\label{def:match}
	Given a KG $\mathcal{G} = \{\mathcal{E}, \mathcal{R}, \mathcal{F}\}$ and an \textit{instruction} (graph pattern) $\tilde{\mathcal{G}} = (\tilde{\mathcal{E}}, \tilde{\mathcal{R}}, \tilde{\mathcal{F}})$, a match $h$ of $\tilde{\mathcal{G}}$ in $\mathcal{G}$ is a mapping from $\mathsf{Const} \cup \mathsf{Var}$ to $\mathsf{Const}$ such that the mapping $h$ maps constants to themselves and variables to constants; if the image of $\tilde{\mathcal{G}}$ under $h$ is contained within $\mathcal{G}$, then $h$ is a match. 
\end{definition}

Additionally, Basic Graph Pattern (BGP) could be augmented with relational operations, including \textit{projection}, \textit{join}, \textit{union}, \textit{difference}, \textit{optional}, and \textit{filter}. Those graph patterns are defined as Complex Graph Pattern (CGP). We summary various graph patterns, operations, and related literature in Table \ref{tab:GPM}.

\begin{table}[t]
\centering
\caption{Summary of graph pattern matching queries.}
\begin{tabular}{ccc}
  \toprule
  Type & Operations & Related Works\\
  \midrule
  BGP & Matching &\cite{martinez1997algorithm} \cite{Fan2012GraphPM} \cite{Chen2018OnEO} \cite{Cheng2008FastGP}\\
  \midrule
  \multirow{6}{*}{CGP} & Projection & \cite{Buzmakov2015RevisitingPS} \\
  ~ & Join & \cite{Fuchs2020EdgeFrameWO} \cite{Vidal2010EfficientlyJG}\\
  ~ & Union & \cite{Shoudai2018PolynomialTL} \\
  ~ & Difference & \cite{ferre2018sparql} \\
  ~ & Optional & \cite{Mennicke2019FastDS} \\
  ~ & Filter & \cite{ferre2018sparql} \\
  \bottomrule
\end{tabular}
\label{tab:GPM}
\end{table}

\nosection{Graph Navigation (GN)} 
While graph pattern matching provides most of the query functionality, it is also helpful to support navigation towards the graph topology. Graph navigation has been widely studied by the research community \cite{Wood2012QueryLF, Barcel2013QueryingGD} and is adopted in modern graph query languages such as Gremlin \cite{rodriguez2015gremlin}. One typical example of such a query is `finding all friends-of-a-friend of some person' in a social network. Here we are not only interested in the immediate acquaintances of a person, but also the people she might know through other people; namely, her friends-of-a-friend, their friends, and so on. Traditionally, \textit{Path Query} is a basic component of GN functionality, where path query navigates through arbitrary number of edges in the graph. In particular, unlike solutions to GPM that have a fixed-arity output, paths do not have a fixed-arity, therefore we cannot directly define a mapping from variables to constants as in the case of GPM.

\begin{definition}[Path Query]\label{definition:pathquery}
    Given a KG $\mathcal{G} = \{\mathcal{E}, \mathcal{R}, \mathcal{F}\}$, a path query is defined in the form of $P=h\to^\gamma t$, where $h, t$ specify the beginning (head) and the ending (tail) entities of the path, and $\gamma$ denotes the traversal condition on the paths.
\end{definition}

In a more complex and popular form, condition $\gamma$ could also be expressed as a regular expression. Those path queries with regular expressions are called Regular Path Queries (RPQs). And two-way Regular Path Queries (2RPQ) further allows the inverse on traversal, i.e., traverse in the backwards direction. Additionally, Conjunctive Regular Path Queries (CRPQ) supports complex conjunctions of path queries. 
We summary these categories of GN queries and their related literature in Table \ref{tab:GN}.

\begin{table}[t]
\centering
\caption{Summary of graph navigation queries.}
\begin{tabular}{ccc}
  \toprule
  Categories & Features & Related Works\\
  \midrule
  RPQ & Regular expressions & \cite{Calvanese2003ReasoningOR}\\
  2RPQ & Regular expressions with inverse & \cite{Calvanese1999RewritingOR} \cite{Calvanese2003ReasoningOR}\\
  CRPQ & Conjunctions of RPQ & \cite{Consens1990GraphLogAV} \cite{Sasaki2020StructuralIF}\\
  \bottomrule
\end{tabular}
\label{tab:GN}
\end{table}

\subsection{Problem Description of Privacy Preserving KG Query}
Privacy preserving KG query addresses the problem of query over securely-stored graph databases. Recall in Section \ref{sec:merge:storage}, the graph database is distributed and stored between two parties after merging. Each party holds its own triple table and property table, where the properties of the common entities are stored in secret share format and others are stored in plaintext. Informally, we defined the privacy-preserving KG query problem as follows:

\begin{definition}[Privacy preserving KG query]
Assume there are $n$ parties, each party holds a private KG: $\mathcal{G}_i=\{\mathcal{E}_i, \mathcal{R}_i, \mathcal{F}_i\}$, and each party also has a property set $\mathcal{X}_i$ describing its entity set $\mathcal{E}_i$ which is stored in secret sharing format. 
For a general query, the purpose of privacy preserving KG query is find the matched result in all the KGs, on the basis of protecting KG privacy. 
\end{definition}

Intuitively, this problem seems to have a close relationship with the widely-studied \emph{Protected Database Search} (PDS) \cite{song2000practical} problem. 
In the literature, the problem of PDS has been widely-studied from different angles, including Private Information Retrieval (PIR) \cite{Kiayias2015OptimalRP, Canetti2017TowardsDE, Boyle2017CanWA, Patel2018PrivateSI, Angel2018PIRWC, Ali2019CommunicationComputationTI}, Searchable Symmetric Encryption (SSE) \cite{cash2013highly}, Order-Preserving Encryption (OPE) \cite{ahmed2019semi}. Nevertheless, most of the existing works focus on relational databases or NoSQL databases. 
There is quite limited study on searching over protected graph databases. Unlike relational databases, graph databases by design maintain the relationship between nodes, which allow fast and efficient connection check between two nodes \cite{besta2019demystifying,cui2020highly}.
The most related work on protected graph database query is GraphSE$^2$ \cite{lai2019graphse2}, which addresses the protected graph query problem by leveraging a SSE scheme called Oblivious Cross-Tags protocol (OXTs) \cite{cash2013highly}. 
However, GraphSE$^2$ inherits the leakage from the original OXT protocol and inevitably leaks access patterns. Therefore, GraphSE$^2$ is not an ideal solution for privacy preserving KG query. 
To date, there is few research on solving privacy preserving KG query problem securely and efficiency.

\subsection{Possible Solutions for Privacy Preserving KG Query}
One possible solution for general-purpose privacy preserving KG query system is to adopt the idea from industrial systems, Gremlin \cite{rodriguez2015gremlin}, SPARQL \cite{prudhommeaux2008sparql} and Cypher \cite{francis2018cypher}. In this section, we introduce the idea from \cite{cryptoeprint:2020:1415}, and describe how to adopt implement a privacy-preserving graph traversal machine. 

To begin with, Gremlin uses traverser with instructions to allow the navigation towards graph topology and the graph pattern matching. With the two most fundamental query functionalities of GPM and GN, we use the term ``\textit{instruction}'' to represent a single atomic operation for GPM or GN functionality and also use the term ``\textit{multi-instruction}'' as a conjunction of single instructions. Additionally, we take a simplified formalization of a graph traversal machine, that is,
\begin{equation*}
    T=(U\times\Psi),
\end{equation*}
where $U\subseteq\mathcal{E}\cup \mathcal{R}\cup \mathcal{F}$ is the traverser's location set and $\Psi$ is the query instruction set. Moreover, we consider the locations for each traversal as sensitive apart from the beginning and the ending  of the traversal, since they are the input and output of the traversal. Informally, we denote the traverser for our privacy-preserving graph traversal machine as $T=(\langle U\rangle, \Psi)$.





\begin{figure}[t]
\centering
\tikzset{every picture/.style={line width=0.8pt}} 

\begin{tikzpicture}[x=0.75pt,y=0.75pt,yscale=-0.75,xscale=0.75]

\draw   (10,130) -- (90,130) -- (90,150) -- (10,150) -- cycle ;
\draw   (10,150) -- (90,150) -- (90,170) -- (10,170) -- cycle ;
\draw   (83,68) -- (163,68) -- (163,98) -- (83,98) -- cycle ;
\draw    (163,83) -- (200,83) ;
\draw [shift={(203,83)}, rotate = 180] [fill={rgb, 255:red, 0; green, 0; blue, 0 }  ][line width=0.08]  [draw opacity=0] (10.72,-5.15) -- (0,0) -- (10.72,5.15) -- (7.12,0) -- cycle    ;
\draw    (50,130) .. controls (41.85,109.8) and (45,83.99) .. (80.23,83.03) ;
\draw [shift={(83,83)}, rotate = 180.5] [fill={rgb, 255:red, 0; green, 0; blue, 0 }  ][line width=0.08]  [draw opacity=0] (10.72,-5.15) -- (0,0) -- (10.72,5.15) -- (7.12,0) -- cycle    ;
\draw   (195,130) -- (275,130) -- (275,150) -- (195,150) -- cycle ;
\draw   (195,150) -- (275,150) -- (275,170) -- (195,170) -- cycle ;
\draw    (235,60) -- (235,127) ;
\draw [shift={(235,130)}, rotate = 270] [fill={rgb, 255:red, 0; green, 0; blue, 0 }  ][line width=0.08]  [draw opacity=0] (10.72,-5.15) -- (0,0) -- (10.72,5.15) -- (7.12,0) -- cycle    ;
\draw  [dash pattern={on 0.84pt off 2.51pt}] (10,10) -- (280,10) -- (280,110) -- (10,110) -- cycle ;
\draw   (195,40) -- (275,40) -- (275,60) -- (195,60) -- cycle ;

\draw (85,75) node [anchor=north west][inner sep=0.75pt]   [align=left] {$\mathsf{SExecute}$};
\draw (204,74) node [anchor=north west][inner sep=0.75pt]  [font=\normalsize] [align=left] {$\displaystyle \langle \mathbf{b}\rangle $};
\draw (34.92,131) node [anchor=north west][inner sep=0.75pt]  [font=\normalsize,rotate=-359.75] [align=left] {$\displaystyle \langle U\rangle $};
\draw (13.01,149.99) node [anchor=north west][inner sep=0.75pt]  [font=\normalsize,rotate=-359.75] [align=left] {$\displaystyle \psi _{1} ,..,\psi _{|\Psi|}$};
\draw (218.01,130.09) node [anchor=north west][inner sep=0.75pt]  [font=\normalsize,rotate=-359.75] [align=left] {$\displaystyle \langle U^{\prime } \rangle $};
\draw (198.01,149.99) node [anchor=north west][inner sep=0.75pt]  [font=\normalsize,rotate=-359.75] [align=left] {$\displaystyle \psi _{2} ,..,\psi_{|\Psi|} $};
\draw (237,81) node [anchor=north west][inner sep=0.75pt]   [align=left] {$\mathsf{Filter}$};
\draw (12,13) node [anchor=north west][inner sep=0.75pt]   [align=left] {{Single Instruction Evaluation (SIE)}};

\draw (194,42) node [anchor=north west][inner sep=0.75pt]   [align=left] {$\displaystyle \mathcal{E}\cup\mathcal{R}\cup\mathcal{F}$};

\end{tikzpicture}
 
\caption{A framework for secure single instruction query.}
\label{fig:single-query}
\end{figure}
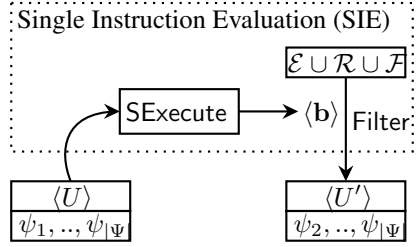

\nosection{Secure single instruction query} On a high-level, single instruction query takes a shared location set $\langle U\rangle$ and an instruction $\psi \in \{\psi _{1} ,...,\psi _{|\Psi|}\}$ as the input of a traverser $t$, and outputs a new shared location set $\langle U^\prime\rangle$. 
Specifically, this procedure can be divided into two main steps, i.e., $\mathsf{SExecute}$ and $\mathsf{Filter}$, as is shown in Figure \ref{fig:single-query}. 
$\mathsf{SExecute}$ is an atomic instruction processing operation, which inputs $\langle U\rangle$ and $\psi$ and further outputs a shared binary vector $\langle \mathbf{b} \rangle$ with its length equal to $|\mathcal{E}\cup \mathcal{R}\cup \mathcal{F}|$. Here, the shared binary vector $\langle \mathbf{b} \rangle$ acts as an indication vector. 
After it, given the shared binary vector $\langle \mathbf{b} \rangle$ and the traverser's entire location set $\mathcal{E}\cup \mathcal{R}\cup \mathcal{F}$, $\mathsf{Filter}$ gets a new shared location set $\langle U^\prime\rangle$ by multiplying $\langle \mathbf{b} \rangle$ and $\mathcal{E}\cup \mathcal{R}\cup \mathcal{F} $. 

Note that the detailed construction of $\mathsf{SExecute}$ varies with KG storage methods, KG graph models, and even with underlying graph data. Therefore, in this paper, we will not go deeper into the detail of $\mathsf{SExecute}$'s construction and only give an abstraction of $\mathsf{SExecute}$.

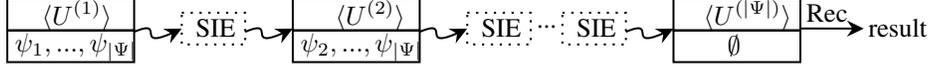
\begin{figure*}[t]   
\centering
\tikzset{every picture/.style={line width=0.75pt}} 

\begin{tikzpicture}[x=0.75pt,y=0.75pt,yscale=-0.8,xscale=0.8]

\draw   (10,11.14) -- (90,11.14) -- (90,31.14) -- (10,31.14) -- cycle ;
\draw   (10,31.14) -- (90,31.14) -- (90,51.14) -- (10,51.14) -- cycle ;
\draw  [dash pattern={on 0.84pt off 2.51pt}] (120,20) -- (160,20) -- (160,40) -- (120,40) -- cycle ;
\draw    (90,32.07) .. controls (111.61,22.81) and (87.44,45.48) .. (117.57,31.17) ;
\draw [shift={(120,30)}, rotate = 513.77] [fill={rgb, 255:red, 0; green, 0; blue, 0 }  ][line width=0.08]  [draw opacity=0] (10.72,-5.15) -- (0,0) -- (10.72,5.15) -- (7.12,0) -- cycle    ;
\draw    (160,32.07) .. controls (181.61,22.81) and (157.44,45.48) .. (187.57,31.17) ;
\draw [shift={(190,30)}, rotate = 513.77] [fill={rgb, 255:red, 0; green, 0; blue, 0 }  ][line width=0.08]  [draw opacity=0] (10.72,-5.15) -- (0,0) -- (10.72,5.15) -- (7.12,0) -- cycle    ;
\draw    (510,29.93) -- (547,29.99) ;
\draw [shift={(550,30)}, rotate = 180.1] [fill={rgb, 255:red, 0; green, 0; blue, 0 }  ][line width=0.08]  [draw opacity=0] (10.72,-5.15) -- (0,0) -- (10.72,5.15) -- (7.12,0) -- cycle    ;
\draw    (400,32.07) .. controls (421.61,22.81) and (397.44,45.48) .. (427.57,31.17) ;
\draw [shift={(430,30)}, rotate = 513.77] [fill={rgb, 255:red, 0; green, 0; blue, 0 }  ][line width=0.08]  [draw opacity=0] (10.72,-5.15) -- (0,0) -- (10.72,5.15) -- (7.12,0) -- cycle    ;
\draw   (190,11.14) -- (270,11.14) -- (270,31.14) -- (190,31.14) -- cycle ;
\draw   (190,31.14) -- (270,31.14) -- (270,51.14) -- (190,51.14) -- cycle ;
\draw    (270,32.07) .. controls (291.61,22.81) and (267.44,45.48) .. (297.57,31.17) ;
\draw [shift={(300,30)}, rotate = 513.77] [fill={rgb, 255:red, 0; green, 0; blue, 0 }  ][line width=0.08]  [draw opacity=0] (10.72,-5.15) -- (0,0) -- (10.72,5.15) -- (7.12,0) -- cycle    ;
\draw  [dash pattern={on 0.84pt off 2.51pt}] (300,20) -- (340,20) -- (340,40) -- (300,40) -- cycle ;
\draw  [dash pattern={on 0.84pt off 2.51pt}] (360,20) -- (400,20) -- (400,40) -- (360,40) -- cycle ;
\draw   (430,11.14) -- (510,11.14) -- (510,31.14) -- (430,31.14) -- cycle ;
\draw   (430,31.14) -- (510,31.14) -- (510,51.14) -- (430,51.14) -- cycle ;

\draw (30,10) node [anchor=north west][inner sep=0.75pt]  [font=\normalsize,rotate=-359.75] [align=left] {$\displaystyle \langle U^{(1)}\rangle $};
\draw (13.01,31.14) node [anchor=north west][inner sep=0.75pt]  [font=\normalsize,rotate=-359.75] [align=left] {$\displaystyle \psi _{1} ,...,\psi_{|\Psi|} $};
\draw (127,22) node [anchor=north west][inner sep=0.75pt]   [align=left] {SIE};
\draw (342,26) node [anchor=north west][inner sep=0.75pt]   [align=left] {...};
\draw (447,10) node [anchor=north west][inner sep=0.75pt]  [font=\normalsize,rotate=-359.75] [align=left] {$\displaystyle \langle U^{(|\Psi|)} \rangle $};
\draw (462.11,32.08) node [anchor=north west][inner sep=0.75pt]  [font=\normalsize,rotate=-359.75] [align=left] {$\displaystyle \emptyset $};
\draw (551,22) node [anchor=north west][inner sep=0.75pt]   [align=left] {result};
\draw (512,12.93) node [anchor=north west][inner sep=0.75pt]   [align=left] {Rec};
\draw (214.92,10) node [anchor=north west][inner sep=0.75pt]  [font=\normalsize,rotate=-359.75] [align=left] {$\displaystyle \langle U^{(2)}\rangle $};
\draw (193.01,31.14) node [anchor=north west][inner sep=0.75pt]  [font=\normalsize,rotate=-359.75] [align=left] {$\displaystyle \psi _{2} ,...,\psi_{|\Psi|} $};
\draw (307,22) node [anchor=north west][inner sep=0.75pt]   [align=left] {SIE};
\draw (367,22) node [anchor=north west][inner sep=0.75pt]   [align=left] {SIE};

\end{tikzpicture}
\caption{A general framework for secure multi-instruction query.}
\label{fig:seq_multi-query}
\end{figure*}

\nosection{Secure multi-instruction query}
Given above secure single instruction query, it is also challenging to compose the single instruction queries to obtain a secure and efficient general framework for multi-instruction query. One native solution is to sequentially perform Single Instruction Evaluation (SIE), and then use secure matrix multiplication of $\langle \mathbf{b}^{(j)} \rangle$ and $\mathcal{E}\cup\mathcal{R}\cup\mathcal{F}$ to get the shared traverser location $\langle U^{(j+1)}\rangle$. As is shown in Figure \ref{fig:seq_multi-query}, the traverer begins with the entire location set $\mathcal{E}\cup\mathcal{R}\cup\mathcal{F}$, and an instruction set $\Psi=\{\psi_1,...,\psi_{|\Psi|}\}$. Then SIE sequentially evaluates each instruction ($\psi_1, \psi_2, ..., \psi_{|\Psi|}$), and receives a location set after every evaluation. Finally, we call a reconstruction on the resulting location $\langle U^{(|\Psi|)}\rangle$ and obtain the final results.


\section{Privacy Preserving KG Representation}
In this section, we first simply summarize the taxonomy of traditional KG representation learning. We then describe the privacy preserving KG representation problem. We finally present possible solutions for the above problem. 

\subsection{Taxonomy of KG Representation}
KG representation learning, aka KG embedding, aims to learn low-dimensional embeddings of entities and relations \cite{ji2020survey}. KG representation learning has been widely studied recently, because it significantly affects the performance of downstream KG completion and application tasks. 
 According to \cite{ji2020survey,wang2017knowledge}, most existing works on KG representation mainly focus on two directions, i.e., \textit{encoding model} and \textit{scoring function}. The former aims to encode the interactions of entities and relations through specific model architectures, while the later is used to measure the plausibility of facts. Commonly used encoding models include linear model, factorization model, neural network (NN) model, Convolutional Neural Network (CNN), Recurrent Neural Network (RNN), and Graph Neural Network (GNN). 
Popularly adopted scoring functions consist of distance based methods and similarity based methods. 
We summarize some related work of KG representation in Table \ref{tab:kg-repre}. More details can be found in \cite{ji2020survey,wang2017knowledge}. 

\begin{table}[t]
\centering
\caption{Summary of KG representation}
\begin{tabular}{ccc}
  \toprule
  Directions & Methods & Related Works\\
  \midrule
  \multirow{6}{*}{Encoding Model} & Linear & \cite{bordes2013translating,wang2018multi} \\
  ~ & Factorization & \cite{nickel2011three,jenatton2012latent} \\
  ~ & NN & \cite{dong2014knowledge,socher2013reasoning} \\
  ~ & CNN & \cite{shang2019end,dettmers2018convolutional} \\
  ~ & RNN & \cite{gardner2014incorporating,neelakantan2015compositional} \\
  ~ & GNN & \cite{nathani2019learning,vashishth2019composition,schlichtkrull2018modeling} \\
  \midrule
  \multirow{2}{*}{Scoring Function} & Distance based & \cite{bordes2011learning,bordes2013translating,lin2015learning} \\
  ~ & Similarity based & \cite{xue2018expanding,zhang2019interaction,xu2019relation} \\
  \bottomrule
\end{tabular}
\label{tab:kg-repre}
\end{table}

\subsection{Problem Description of Privacy Preserving KG Representation}
KG representation learning is nontrivial under KG isolation setting, since not only entity properties, but also relations between entities, are securely stored by multi-parties after merging, as are shown in Table \ref{tab:APT} and Table \ref{tab:BPT}. 
Formally, privacy preserving KG representation is defined as follows. 

\begin{definition}[Privacy preserving KG representation]
Given $n$ parties, each of whom has an individual KG, i.e., $\mathcal{G}_i=\{\mathcal{E}_i, \mathcal{R}_i, \mathcal{F}_i\}$, and each party also has a property set $\mathcal{X}_i$ describing its entity set $\mathcal{E}_i$, where $\mathcal{X}_i$ includes the properties ($\tilde{\mathcal{X}}$) of the common entity set ($\tilde{\mathcal{E}}$) that are stored in secret sharing format, and $i\in \{1, 2, ..., n\}$. The purpose of privacy preserving KG representation is to learn low-dimensional entity embedding $\bm\mu ^{i,j}$ for $j$-th entity of party $i$, and relation embedding $\bm\eta ^{i,j,k}$ for the relation between $j$-th entity and $k$-th entity of party $i$, on the basis of protecting each party's private data. 
\end{definition}

So far, there is only limited work on privacy preserving graph embedding and graph neural network \cite{zhou2020privacy,zheng2020asfgnn}. For example, Zhou et al. \cite{zhou2020privacy} proposed a server-aided privacy preserving GNN learning method, which adopts the idea of split learning \cite{vepakomma2018split} and splits the computation graph of GNN into two parts. The private graph data related computations are done by data holders and the rest hidden layer related computations are done by a neutral server. Zheng et al. \cite{zheng2020asfgnn} proposed to combine federated learning with automated machine learning to solve the privacy and data non-independent identically distributed problem in data isolation setting. 
Although both approaches can protect data privacy to a certain extent, they are not provable secure. 

\subsection{Possible Solutions for Privacy Preserving KG Representation Learning}

We divide the solution for privacy preserving KG representation learning into the following three steps based on the main steps in traditional GNN. 

\nosection{Secure initial entity (node) embedding generation}
The first step is generating initial node embeddings securely for multiple parties using their node features. Traditionally, initial node embeddings are generated by using a non-linear transformations, i.e., $\textbf{h}^0 = \sigma (\textbf{x}\textbf{W})$, where $\textbf{x}$ is node feature, $\textbf{W}$ is a weight matrix, and $\sigma$ is a non-linear active function. Under KG isolation setting, after secure KG storage, node features are either kept by a single party or secretly shared by multiply parties, as are shown in Table \ref{tab:APT} and Table \ref{tab:BPT}. Besides, the weight matrix ($\textbf{W}$) should also be kept secretly for privacy concern. Motivated by existing work \cite{demmler2015aby}, we propose to store $\textbf{W}$ in secret sharing format. That is, each party $i$ holds a share $\left\langle \textbf{W} \right\rangle_i$. Therefore, secure initial node embedding generation becomes the following problem: each party $i$ holds its private feature $\textbf{x}$ or its shares $\left\langle \textbf{x} \right\rangle_i$, and the weight share $\left\langle \textbf{W} \right\rangle_i$, and all the parties want to calculate $\textbf{h}^0 = \sigma (\textbf{x}\textbf{W})$ securely and collaboratively, such that each party holds a share $\left\langle \textbf{h}^0 \right\rangle_i$ at the end of this step. Besides, $\sigma$ are either non-linear continuous active functions such as Sigmoid and Tanh which are not cryptography-friendly, or piece-wise active functions such as RELU which rely on time-consuming secure comparison protocols. For the non-linear continuous active functions, existing works propose to use polynomials \cite{aono2016scalable} or piece-wise functions \cite{mohassel2017secureml} to approximate them. 

We now present how to calculate $\textbf{h}^0$ securely for $n$ parties in details. We assume both $\textbf{x}$ and $\textbf{W}$ are securely shared\footnote{Note that $\textbf{x}$ can be easily transformed into secret sharing format, even when it is originally held by a single party.}, and summarize the calculation procedure in Algorithm 1. 
It mainly has three steps. The first step is securely calculating $\textbf{x} \textbf{W}$ for multiple parties, as is shown in line \ref{algo-init-wx-begin}-\ref{algo-init-wx-end}, which involves secret sharing based addition and multiplication. The second step calculates polynomial variables, as is shown in line \ref{algo-init-poly-begin}-\ref{algo-init-poly-end}. The last step approximates $\textbf{h}^0$ by using polynomial, as is shown in line \ref{algo-init-hidden-begin}-\ref{algo-init-hidden-end}. 
Here, we use a two-order polynomial as an example to approximate the non-linear functions as follows
\begin{equation}
\textbf{h}^0 = \frac{1}{1 + e^{-\textbf{z}}} \approx q_0 + q_1 \textbf{z} + q_2 \textbf{z}^2,
\end{equation}
where the coefficients could be set using different methods \cite{chen2020homomorphic}. After this step, each party holds the same number of entities, with the same dimension of embeddings. 

\begin{algorithm}[th]
\caption{Securely  generate initial node embeddings for $n$ parties}\label{algo-init-hidden}
\KwIn {Public polynomial coefficients ($q_0, q_1, q_2$), for $\forall i\in \{1, 2, ..., n\}$, party $i$ holds its shares $\left\langle \textbf{x} \right\rangle_i$ and $\left\langle \textbf{W} \right\rangle_i$}
\KwOut{For $\forall i\in \{1, 2, ..., n\}$, party $i$ gets its share $\left\langle \textbf{h}^0 \right\rangle_i$}
\underline{\textbf{Securely calculate $\textbf{z} = \textbf{x}\textbf{W}$:}}\\
\For{$i = 1$ to $n$}
{\label{algo-init-wx-begin}
Party $i$ locally calculates $\left\langle \textbf{x} \right\rangle_i  \left\langle \textbf{W} \right\rangle_i$ as $i$-share\\
    \For{$j = 1$ to $n$ and $j \ne i$}
    {
    Party $i$ and party $j$ securely calculate $\left\langle \textbf{x} \right\rangle_i \left\langle \textbf{W} \right\rangle_j$ using secret sharing \textbf{MUL} primitive, and after that party $i$ gets its $i$-share $\left\langle \left\langle \textbf{x} \right\rangle_i \left\langle \textbf{W} \right\rangle_j \right\rangle_i$ and party $j$ gets its $j$-share $\left\langle \left\langle \textbf{x} \right\rangle_i \left\langle \textbf{W} \right\rangle_j \right\rangle_j$\\
    }
}
\For{$i = 1$ to $n$}
{
Party $i$ locally calculates the summation of all i-shares and denote it as $\left\langle \textbf{z} \right\rangle_i$\\
}
\label{algo-init-wx-end}
\underline{\textbf{Securely calculate $\textbf{z}^2$:}}\\
\For{$i = 1$ to $n$}
{\label{algo-init-poly-begin}
Party $i$ locally calculates the element-wise square of $\left\langle \textbf{z} \right\rangle_i$ as its $i$-share $\left\langle \textbf{z} \right\rangle_i \odot \left\langle \textbf{z} \right\rangle_i$\\
\For{$j = 1$ to $n$ and $j \ne i$}
    {
    Party $i$ and party $j$ securely calculate $2\left\langle \textbf{z} \right\rangle_i \odot \left\langle \textbf{z} \right\rangle_j$ using secret sharing \textbf{MUL} primitive, and after that party $i$ gets its $i$-share $2\left\langle \left\langle \textbf{z} \right\rangle_i \odot \left\langle \textbf{z} \right\rangle_i \right\rangle_i$ and party $j$ gets its $j$-share $2\left\langle \left\langle \textbf{z} \right\rangle_i \odot \left\langle \textbf{z} \right\rangle_i \right\rangle_j$\\
    }
}
\For{$i = 1$ to $n$}
{
Party $i$ locally calculates the summation of all $i$-shares and denote it as $\left\langle \textbf{z}^2 \right\rangle_i$\\
}\label{algo-init-poly-end}
\underline{\textbf{Securely calculate $\textbf{h}^0$:}}\\
\For{$i = 1$ to $n$}
{\label{algo-init-hidden-begin}
Party $i$ locally calculates $\left\langle \textbf{h}^0 \right\rangle_i = q_0 / n + q_1 \left\langle \textbf{z} \right\rangle_i + q_2 \left\langle \textbf{z}^2 \right\rangle_i$\\
}\label{algo-init-hidden-end}
\end{algorithm}

\begin{figure}
\centering
\includegraphics[width=6cm]{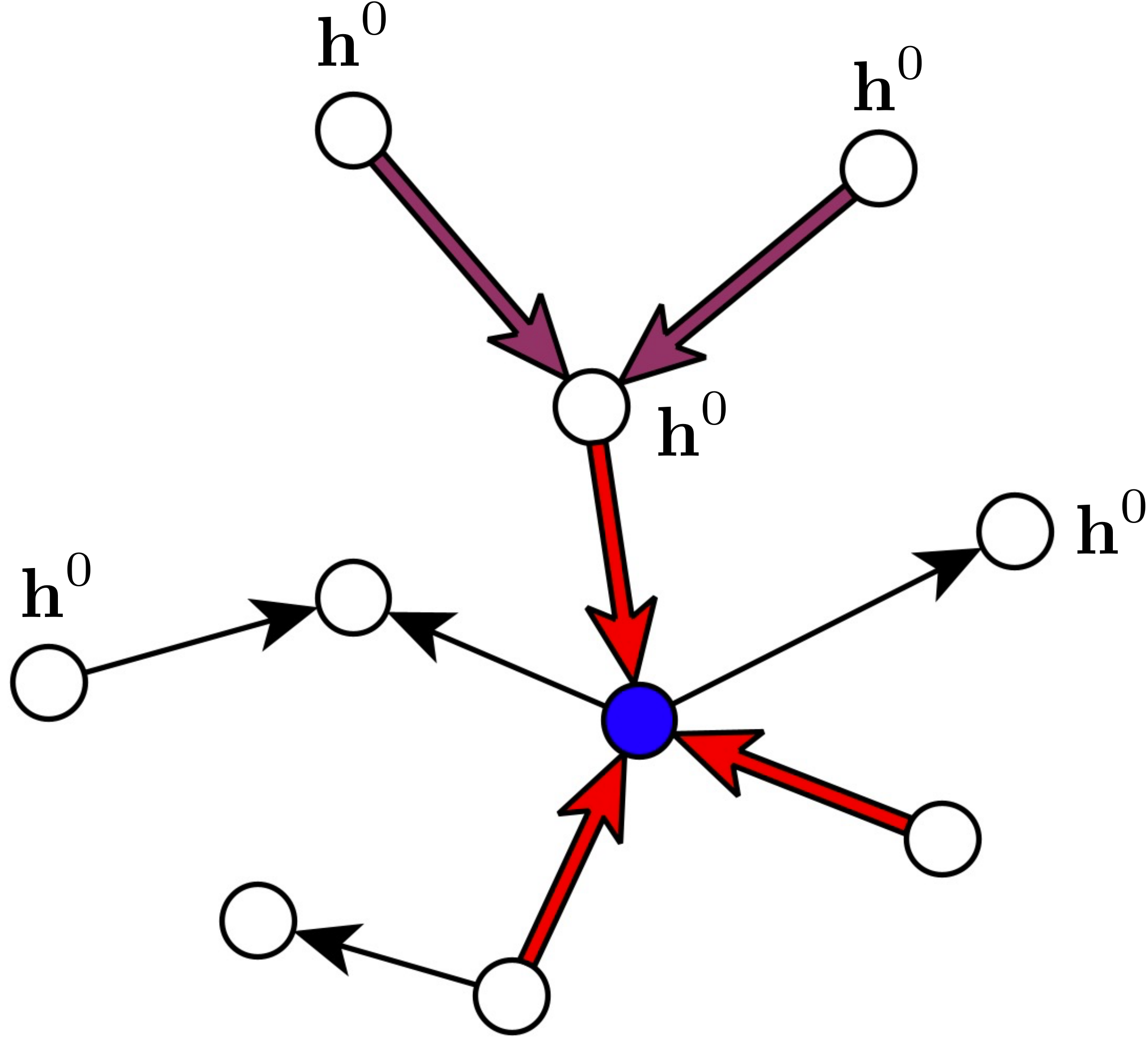}
\caption{Traditional embedding propagation on KG. }
\label{fig:represent_message}
\end{figure}

\nosection{Secure embedding propagation }
The second step is propagating node embeddings (aka. passing message) securely for multiple parties using their initial node embeddings and relations between nodes on KGs, as is shown in Figure \ref{fig:represent_message}. Existing works on GNN have proposed different kinds of embedding propagation methods, e.g., convolution based \cite{hamilton2017inductive}, attention based \cite{velivckovic2017graph}, and gated mechanism based \cite{li2015gated}, and their mixtures \cite{liu2019geniepath}. Take GraphSAGE \cite{velivckovic2017graph}---a classic convolution based GNN---for example, it first aggregates neighborhood embeddings and then transforms it using a fully-connected layer, 
\begin{equation}
\textbf{h}_{\mathcal{N}(v)}^k \leftarrow \text{AGG}_k(\{\textbf{h}_u^{k-1}, \forall u \in \mathcal{N}(v)\}),
\end{equation}
\begin{equation}
\textbf{h}^k_v \leftarrow (\textbf{W}^k \cdot \text{CONCAT} ( \textbf{h}^{k-1}_v,\textbf{h}_{\mathcal{N}(v)}^k ) ),
\end{equation}
where the aggregator functions $\text{AGG}_k$ are of three types, i.e., Mean, LSTM, and Pooling. Under KG isolation setting, secure embedding propagation becomes challenging because initial node embeddings are shared by all the parties, and relations between nodes are also separated by all the parties. 

We present how to perform embedding propagation securely under KG isolation setting in Algorithm \ref{algo-message-pass}, where we take Mean aggregator and Pooling aggregator for example and leave other aggregators as future work. 
Before secure embedding propagation, each party holds its own graph $\mathcal{N}^i$ and the shares of initial node embedding $\left\langle \textbf{h}^0 \right\rangle_i$. Besides, the weight matrices of $k$-th layer $\textbf{W}^k$ should also be kept secretly for privacy concern. 
In Algorithm \ref{algo-message-pass}, line \ref{algo-message-pass-mean-begin}-\ref{algo-message-pass-mean-end} shows how embeddings are propagated using MEAN aggregator. Specifically, all the parties first locally calculate the mean of their shares, as is shown in line \ref{algo-message-pass-mean-indivi-begin}-\ref{algo-message-pass-mean-indivi-end}. They then securely calculate the mean of embeddings using secret sharing division protocol, as is shown in line \ref{algo-message-pass-mean-indivi-mid}. 
Alternatively, one can choose Pooling aggregator to do embedding propagation, as is shown in line \ref{algo-message-pass-pool-begin}-\ref{algo-message-pass-pool-end} in Algorithm \ref{algo-message-pass}. 
For Pooling aggregator, we apply an element-wise max-pooling operation to aggregate information across neighbors, using secret sharing ArgMax protocol. 
After it, all the parties concat their local shares (line \ref{algo-message-pass-concat}) and calculate a non-linear transformation using Algorithm \ref{algo-init-hidden}. Finally, each party holds a share of entity embedding after $K$-depth of propagation. 

\begin{algorithm}[th]
\caption{Secure embedding propagation for $n$ parties}\label{algo-message-pass}
\KwIn {Embedding propagation depth $K$, for $\forall i\in \{1, 2, ..., n\}$, party $i$ holds its graphs $\mathcal{N}^i$, the shares of initial node embedding $\left\langle \textbf{h}^0 \right\rangle_i$, and the shares of $K$ weight matrices $\left\langle \textbf{W}^k \right\rangle_i$}
\KwOut{For $\forall i\in \{1, 2, ..., n\}$, party $i$ gets its share of node embeddings after embedding propagation $\left\langle \textbf{h}^k \right\rangle_i$}
\For{$k=1$ to $K$}
{
    \underline{\textbf{For secure Mean aggregator:}}\\
    \For{$v \in \mathcal{V}$}
    {\label{algo-message-pass-mean-begin}
        \For{$i = 1$ to $n$ in parallel}
        {\label{algo-message-pass-mean-indivi-begin}
            Party $i$ locally calculates $\left\langle \textbf{h}_{\mathcal{N}^i(v)}^k \right\rangle_i \leftarrow$ Mean$\left(\left\{ \left\langle \textbf{h}_u^{k-1} \right\rangle_i, \forall u \in \mathcal{N}^i(v) \right\}\right)$ using secret sharing \textbf{LINEAR} primitive \label{algo-message-pass-mean-indivi-mid}\\
        }\label{algo-message-pass-mean-indivi-end}
        Party $\{1, 2, ..., n\}$ securely calculate $\frac{\sum\limits_{i=1}^{n}{\left\langle \textbf{h}_{\mathcal{N}^i(v)}^k\right\rangle_i}}{\sum\limits_{i=1}^{n}{|\mathcal{N}^i(v)|}}$ using secret sharing \textbf{DIV} primitive, and after that each party holds a share of the aggregated result $\left\langle \textbf{h}_{\mathcal{N}(v)}^k \right\rangle_i$ \\
    }\label{algo-message-pass-mean-end}
    \underline{\textbf{For secure Pooling aggregator:}}\\
    \For{$v \in \mathcal{V}$}
    {\label{algo-message-pass-pool-begin}
        Party $\{1, 2, ..., n\}$ securely calculate $\left\langle \textbf{h}_{\mathcal{N}^i(v)}^k \right\rangle_i \leftarrow$ Max-Pooling$\left(\left\{ \left\langle \textbf{h}_u^{k-1} \right\rangle_i, \forall u \in \mathcal{N}^i(v) \right\}\right)$ using secret sharing \textbf{ARGMAX} primitive, and after that each party holds a share of the aggregated result $\left\langle \textbf{h}_{\mathcal{N}(v)}^k \right\rangle_i$ \\
    }\label{algo-message-pass-pool-end}
    \underline{\textbf{For both secure Mean and Pooling aggregators:}}\\
    \For{$v \in \mathcal{V}$}
    {
        \For{$i = 1$ to $n$ in parallel}
        {
            Party $i$ locally calculates $\left\langle \tilde{\textbf{h}}^k \right\rangle_i=$ CONCAT$\left( \left\langle \textbf{h}_{v}^{k-1} \right\rangle_i, \left\langle \textbf{h}_{\mathcal{N}(v)}^k \right\rangle_i \right)$ \label{algo-message-pass-concat}\\
        }
        Party $\{1, 2, ..., n\}$ securely calculate $\sigma (\tilde{\textbf{h}}^k \textbf{W}^k)$ using Algorithm \ref{algo-init-hidden} by taking $\tilde{\textbf{h}}^k$ as $\textbf{x}$ and $\textbf{W}^k$ as $\textbf{W}$, and after that each party holds a share of the result $\left\langle \textbf{h}^k \right\rangle_i$\\
    }
}
\end{algorithm}

\nosection{Secure loss computation}
The third step is computing loss securely for multiple parties using their propagated entity embeddings based on certain tasks. For example, one usually uses cross-entropy loss for classification tasks \cite{hamilton2017inductive} and noise contrastive estimation loss for unsupervised tasks \cite{mikolov2013distributed}. In this step, each party holds a share of entity embedding which can be taken as features, and all the parties want to calculate loss securely together. Existing secure machine learning models \cite{mohassel2017secureml,demmler2015aby} can be directly applied to solve this problem. 
\section{Privacy Preserving KG Completion}
In this section, we first simply summarize the taxonomy of traditional KG representation learning. We then describe the privacy preserving KG representation problem. We finally present possible solutions for the above problem. 


\subsection{Taxonomy of KG Completion}
KG completion, aka KG reasoning, is an important research problem due to the nature of incompleteness of KG. KG completion aims to infer the missing properties and triples. 
According to \cite{ji2020survey}, 
existing works on traditional KG completion are mainly in three types, i.e., embedding-based methods \cite{guan2018shared,shi2018open}, relation path inference \cite{lao2010relational,gardner2014incorporating}, and rule-based reasoning \cite{omran2019embedding,guo2016jointly}. Among them, embedding-based methods are popularly used due to its high efficiency. 
Take triple prediction for example, embedding-based methods first calculate pair-wise scores of all the candidate entities given a target entity, and then rank the top candidate entities to link edges with the target entity.

\subsection{Problem Description of Privacy Preserving KG Completion}

Under KG isolation setting, private KG completion is defined as follows. 

\begin{definition}[Privacy preserving KG completion]
Given $n$ parties, each of whom has an individual KG, i.e., $\mathcal{G}_i=\{\mathcal{E}_i, \mathcal{R}_i, \mathcal{F}_i\}$, and each party also has a sparse property set $\mathcal{X}_i$ describing its entity set $\mathcal{E}_i$. The purpose of privacy preserving KG completion is to infer the missing properties $\mathcal{P}_i=\{x_i|x_i \notin \mathcal{X}\}$ and the missing triples $\mathcal{T}_i=\{(h_i, r_i, t_i)|(h_i, r_i, t_i) \notin \mathcal{F}_i\}$ for any party $i\in \{1, 2, ..., n\}$. 
\end{definition}

Although traditional KG completion has been extensively studied, to our best knowledge, there is few literature on how to do privacy preserving KG completion under KG isolation setting. 

\subsection{Possible Solutions for Privacy Preserving KG Completion}

We propose an embedding-based privacy preserving KG completion approach, and the key of which is building embedding projection functions, using the learnt KG representation. 

First, for privacy preserving property completion, the embedding projection function maps embedding to property. Here, the embeddings are secretly shared by multi-parties, which can be seen as private features. The properties are either secretly shared by multi-parties or held by a single party which can also be transformed to secret sharing format, determined by whether the corresponding entity is a common entity or not, which can be taken as private labels. That is, $\left\langle \textbf{x} \right\rangle = f(\left\langle \textbf{h} \right\rangle, \textbf{W}_{pro})$, where $\left\langle \textbf{x} \right\rangle$ is the secretly shared property, $\left\langle \textbf{h} \right\rangle$ is the secretly shared entity embedding, and $\textbf{W}_{pro}$ is a tensor for property completion. The function $f$ could be any existing neural layers such as a fully-connected layer or a convolutional layer. 

Second, for privacy preserving triple completion, the embedding projection function maps two entity embeddings to a relation. Here, the entity embeddings are secretly shared by multi-parties, and the relation denotes whether two entities are linked and thus are held by a single party. That is, $r = f(\langle \textbf{h}^h \rangle, \langle \textbf{h}^t \rangle, \textbf{W}_{tri})$, where $\langle \textbf{h}^h \rangle$ and $\left\langle \textbf{h}^t \right\rangle$ are the secretly shared head entity embedding and tail entity embedding, and $\textbf{W}_{tri}$ is a tensor for triple completion. The function $f$ could be any existing functions for triple completion such as ProjE \cite{shi2016proje} and SENN \cite{guan2018shared}. Take ProjE for example, $f(\langle \textbf{h}^h \rangle, \langle \textbf{h}^t \rangle, \textbf{W}_{tri}) = \sigma\left( \textbf{W}_{tri} \sigma(\langle \textbf{h}^h \rangle + \langle \textbf{h}^t \rangle) \right)$. 

\begin{figure}
\centering
\includegraphics[width=10cm]{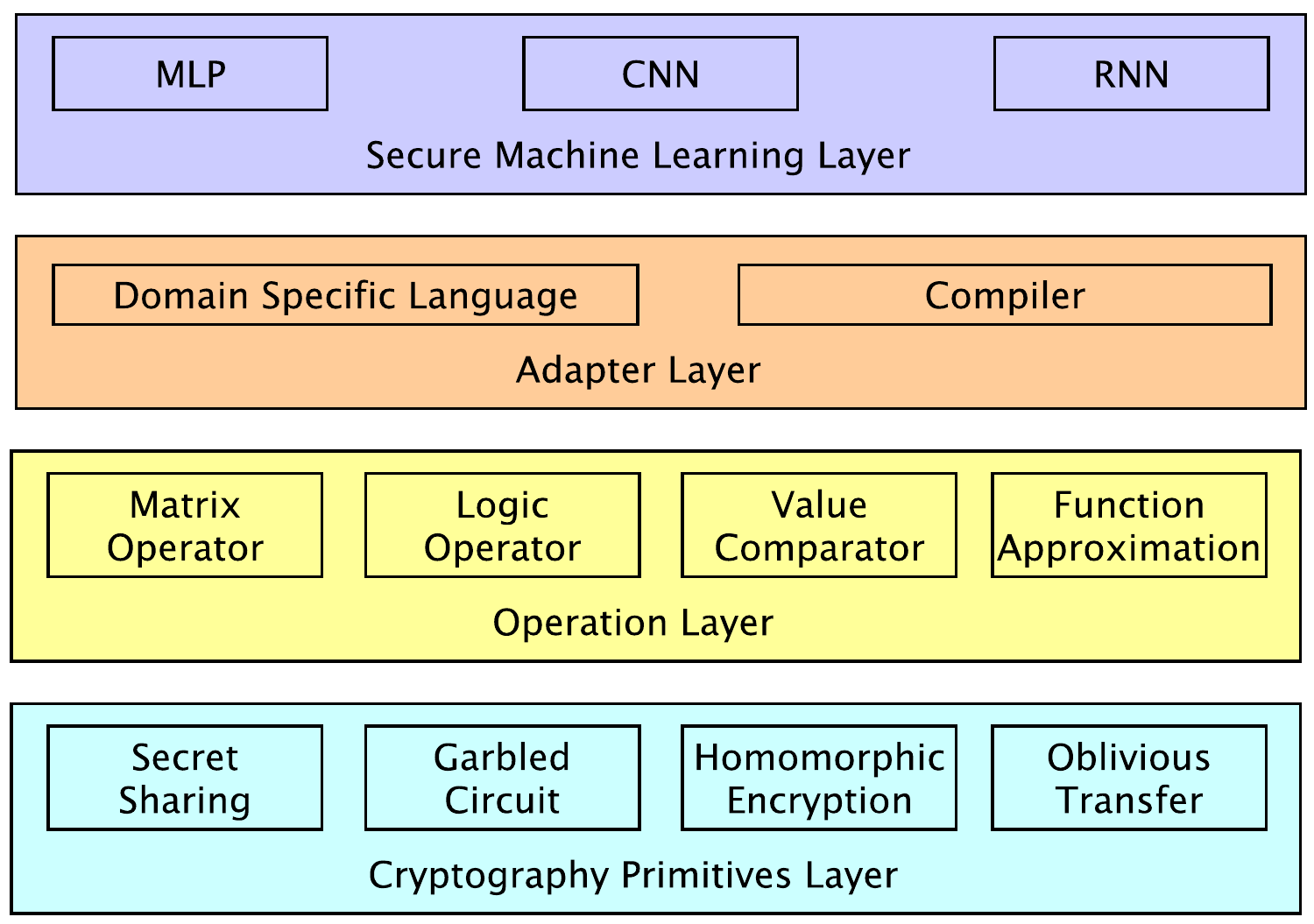}
\caption{Solution for privacy preserving KG completion. }
\label{fig:kg-completion}
\end{figure}

From the above description, we find that the projection functions for both property completion and triple completion need to be changeable due to the variation of different functions. Therefore, the challenge here is how to provide scalable and flexible secure machine learning platforms for one to easily build secure machine learning algorithms to meet the various projection functions in KG completion. This can be solved by developing a system (e.g., Nebula \cite{wuposter}) with hybrid secure computation protocols, rich computation operations, and powerful domain specific language. To this end, we propose a possible solution, as is shown in Figure \ref{fig:kg-completion}, which can be divided into four layers:
\begin{itemize} [leftmargin=*] \setlength{\itemsep}{-\itemsep}
    \item \textbf{Cryptography primitives layer} mainly contains secure computation protocols and their conversions. This is because different protocols have their own advantages and are suitable for different scenarios. 
    \item \textbf{Operation layer} implements popularly used operations in machine learning such as matrix multiplication. Besides, these operations can be performed by various protocols in the cryptography primitives layer. 
    \item \textbf{Adapter layer} defines domain specific language and compiler. This layer aims to  facilitates machine learning developers to develop kinds of algorithms and functions without knowing the complicate cryptographic techniques. 
    \item \textbf{Secure machine learning layer} is built upon the lower-level layers. It provides commonly used secure machine learning models such as Multi-Layer Perception (MLP), to meet the flexible requirements of KG completion tasks. 
\end{itemize}

\section{Privacy Preserving KG-aware Applications}\label{sec:app}

KGs have been applied into various tasks, including question answering \cite{zhang2018variational,huang2019knowledge}, recommender system \cite{cao2019unifying,wang2019kgat}, risk assessment \cite{cheng2019risk}, fraud detection \cite{liu2018heterogeneous,wang2019semi}, and information extraction \cite{hoffmann2011knowledge,koncel2019text}. However, with more people caring privacy and kinds of regulations coming into force, KG isolation becomes a serious problem, which limits the performance KG and prevents KG from being more popularly used. In this section, we present three privacy preserving KG-aware applications and simply describe how can our proposed techniques be applied into these applications. 

\begin{figure}
\centering
\subfigure [Guarantee loan KG of bank $\mathcal{A}$]{ \includegraphics[height=3.2cm]{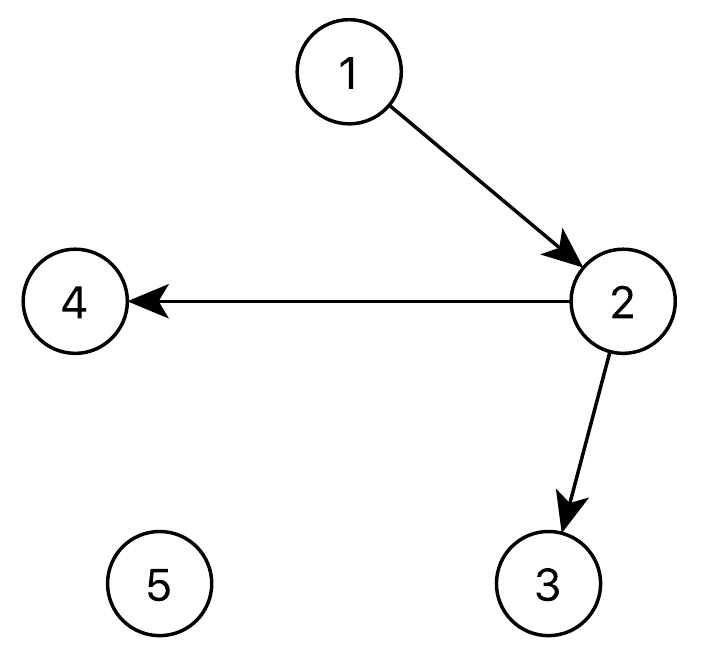}}\hspace{1.0cm}
\subfigure [Guarantee loan KG of bank $\mathcal{B}$] { \includegraphics[height=3.2cm]{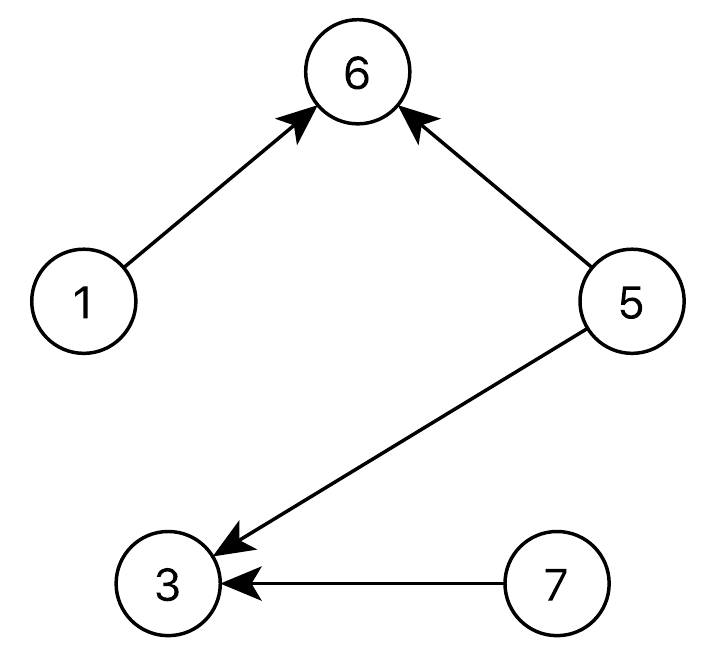}}
\caption{Risk assessment under guarantee loan KG isolation setting, where bank $\mathcal{A}$ and bank $\mathcal{B}$ have three common entities (1, 3, and 5).}
\label{fig:guarantee-loan}
\end{figure}

\subsection{Risk Assessment in Guarantee Loan}
The bank loans are important to the development of small and medium enterprises, and a popularly used way to decrease loan default is guarantee. That is, when a small and medium enterprise needs a bank loan, it can choose another enterprise as a guarantor. As more and more enterprises participate, it becomes a guarantee loan KG. Therefore, how to judge a guarantee is risky or not is important to the risk control in banks. In practice, it is common that enterprises take loans from multiple banks to obtain sufficient funds. 
However, due to privacy considerations,  these banks cannot share customer information with each other, i.e., guarantee loan KGs are isolated, which increases the difficulty of risk control. 

Figure \ref{fig:guarantee-loan} shows a risk assessment example under guarantee loan KG isolation setting. Here, we assume there are two banks and each of them has a guarantee loan KG where node denotes enterprise and edge means guarantee relation. Apparently, the risk assessment ability is limited if only using a single KG. For example, bank $\mathcal{B}$ is likely to allow the guarantee of enterprise 3 to enterprise 1, based its own KG. However, by incorporating the KG of bank $\mathcal{A}$, bank $\mathcal{B}$ will probably refuse the guarantee of enterprise 3 to enterprise 1, because enterprise 1 has guaranteed enterprise 3 indirectly (through enterprise 2) and \textit{guarantee loop} usually has high risk and thus is forbidden. Meanwhile, it is difficult for bank $\mathcal{A}$ to decide whether enterprise 5 is appropriate to guarantee other enterprises since it has no guarantee behaviors in its own KG. By combining the KG of bank $\mathcal{B}$, bank $\mathcal{A}$ will be easier to make the right decision. 

The above described risk assessment problem under guarantee loan KG isolation setting can be solved using our proposed privacy preserving KG query and KG representation methods. First, the guarantee loop detection problem is actually querying whether an enterprise (e.g., 1) has a link to the other enterprise (e.g., 3) in the multiple KGs, on the basis of protecting the KGs. Second, to better assess guarantee loan, privacy preserving KG representation can be used to learning enterprise representations, followed by a binary classification task, similar as \cite{cheng2019risk}. 
Thus, both privacy preserving KG query and KG representation can decrease the risk under guarantee loan KG isolation setting. 

\begin{figure}
\centering
\subfigure [KG of party $\mathcal{A}$]{ \includegraphics[height=3.2cm]{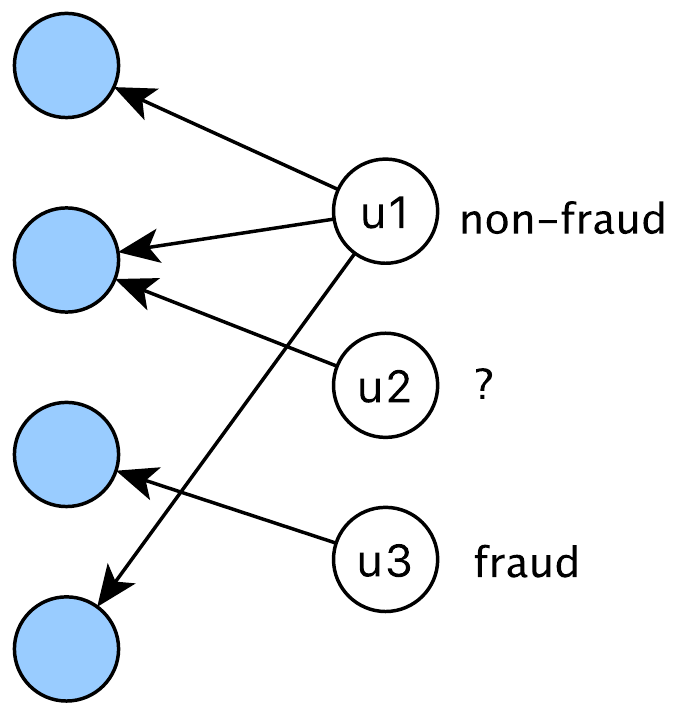}}\hspace{1cm}
\subfigure [KG of party $\mathcal{B}$] { \includegraphics[height=3.2cm]{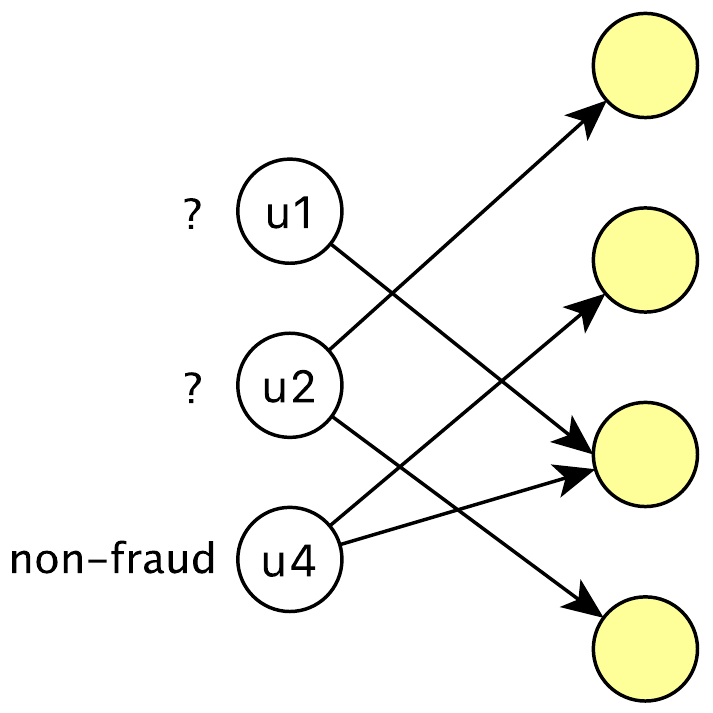}}
\caption{Fraud detection under KG isolation setting.}
\label{fig:fraud-detection}
\end{figure}

\subsection{Fraud Detection}
Fraud detection is a major task for lots of companies, especially for financial companies, due to its direct effect on capital loss. A key to fraud detection is to classify whether a user is a fraud user or non-fraud user \cite{wang2019semi}. 
In practice, users always use the products of multiple companies and each company usually builds its fraud detection system based on its own data. That is, each company has its own KG. It is natural that these companies can improve their fraud detection ability together by combining their KGs. 
Unfortunately, these companies cannot share private user data with each other due to regulation or competition reasons. 
Therefore, the KGs of different companies for fraud detection are isolated.

Figure \ref{fig:fraud-detection} shows a fraud detection example under KG isolation setting. Two companies have different user-item KG, they share some common users ($u1$ and $u2$) but different graph relational data. We assume party $\mathcal{A}$ has a KG with user-device log in data and party $\mathcal{B}$ has another KG with user-wifi log in data. 
Apparently, taking the KGs of both companies into considering will bring more intelligent fraud detection ability. 
The purpose is to build a privacy preserving fraud detection system, i.e., classify whether a user is a fraud user or not, using two KGs. 

The above fraud detection problem under KG isolation setting can by solved by using privacy preserving property completion technique. As we described in Section \ref{sec:intro}, labels can be taken as special properties. With privacy preserving property completion, two companies can build a fraud detection system together while protecting their own KGs. 

\begin{figure}
\centering
\subfigure [KG of party $\mathcal{A}$]{ \includegraphics[height=2.0cm]{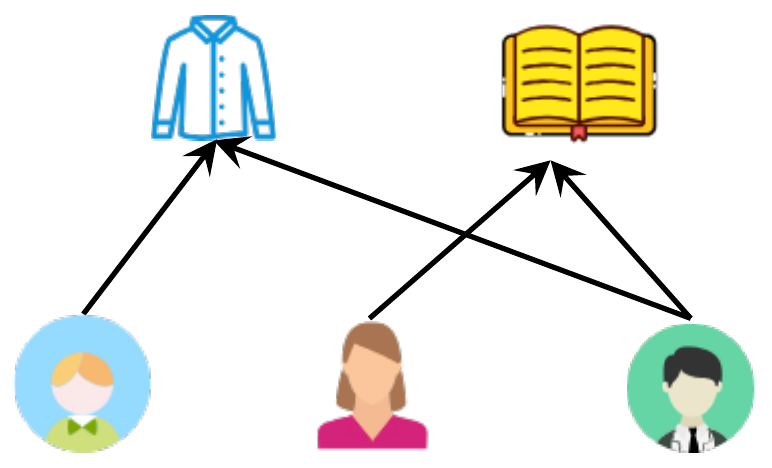}}\label{fig:recsys-a}
\hspace{1cm}
\subfigure [KG of party $\mathcal{B}$] { \includegraphics[height=2.0cm]{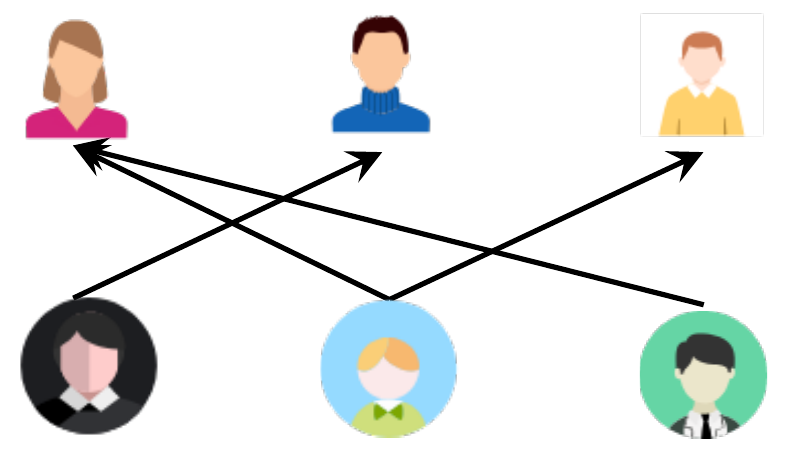}}\label{fig:recsys-b}
\caption{Recommender system under KG isolation setting, where party $\mathcal{A}$ has a user-item interaction KG and party $\mathcal{B}$ has a user social KG.}
\label{fig:recsys}
\end{figure}

\subsection{Recommender System}
Recommender system is a sharp tool to solve the information overload problem, especially when facing the big data era currently. Traditional recommender systems are mainly built based on user-item interaction data, e.g., rating and purchase data. Recent studies have shown that rich side information, e.g., user social information, are effective for improving the recommendation performance \cite{chen2020secure}. However, such information may be isolated by another platform in practice. 
Such data isolation problem limits the performance of recommender systems. 

Figure \ref{fig:recsys} shows a typical case of the KG isolation problem in recommender systems. Figure \ref{fig:recsys} (a) is a KG of party $\mathcal{A}$, which is built using its user-item interaction data. Figure \ref{fig:recsys} (b) is a KG of party $\mathcal{B}$, which is built based on its user social data. This is quite a common situation in reality, since e-commerce platforms such as Amazon have rich user-item interaction data while social media platforms like FaceBook have plenty user social data. How to use the additional data on other platforms to further improve the performance of the recommendation platforms, meanwhile protect the raw data security of both platforms, is a crucial question to be answered \cite{chen2020secure}. 

The above secure recommendation problem under KG isolation setting can be solved by using privacy preserving triple completion technique. Because the essence of recommender system is link prediction problem on KG, i.e., predict whether there is a relation between a user (head entity) to an item (tail entity). With this technique on hand, the development of recommender systems will head to a new direction in the future. 

\section{Conclusion and Future Work}\label{sec:conclusion}
In this paper, we first described the KG isolation problem in practice. We then summarized the open problems in privacy preserving KG, including merging, query, representation, and completion. We formally defined these problems and proposed possible solutions for them. We finally presented three application scenarios of our proposed privacy preserving KG. 
Our work aims to shed light on the future directions in privacy preserving KG under data isolation setting. In the future, we would like to present detailed technical solutions for these problems and further deploy them in real-world applications. 

\bibliographystyle{unsrt}  
\bibliography{references}  

\end{document}